\documentclass[prb,showpacs,twocolumn]{revtex4}

\usepackage{graphics}
\usepackage{epsfig}

\def\vb#1{\mbox{\boldmath$#1$}}
\def\pd#1#2{\frac{\partial #1}{\partial #2}}

\def\wh#1{\widehat{#1}}

\newcommand{\bc}{\begin{center}}
\newcommand{\ec}{\end{center}}
\newcommand{\bt}{\begin{tabbing}}
\newcommand{\et}{\end{tabbing}} 
\newcommand{\be}{\begin{eqnarray*}}
\newcommand{\ee}{\end{eqnarray*}}
\newcommand{\bs}{\begin{slide}}
\newcommand{\es}{\end{slide}}

\begin{document}

\title{A primer on elliptic functions with applications in classical mechanics}

\author{Alain J.~Brizard}
\affiliation{Department of Chemistry and Physics \\ Saint Michael's College, Colchester, VT 05439, USA}

\begin{abstract}
The Jacobi and Weierstrass elliptic functions used to be part of the standard mathematical arsenal of physics students. They appear as solutions of many important problems in classical mechanics: the motion of a planar pendulum (Jacobi), the motion of a force-free asymmetric top (Jacobi), the motion of a spherical pendulum (Weierstrass), and the motion of a heavy symmetric top with one fixed point (Weierstrass). The problem of the planar pendulum, in fact, can be used to construct the general connection between the Jacobi and Weierstrass elliptic functions. The easy access to mathematical software by physics students suggests that they might reappear as useful tools in the undergraduate curriculum.
\end{abstract}

\begin{flushright}
November 26, 2007
\end{flushright}


\maketitle

\section{Introduction}

A long time ago, physics students were well trained in applications of elliptic functions in solving a great variety of problems in classical mechanics. For example, Whittaker's {\it Treatise on the Analytical Dynamics of Particles and Rigid Bodies} \cite{Whittaker} appears to implicitly assume that the reader is fully conversant with the theory of elliptic functions. 

Elliptic functions rapidly fell out the standard physics curriculum over the past fifty years, however, and they are now only mentionned in passing in most standard (modern) textbooks on classical mechanics.\cite{Landau,Taylor,Goldstein} The information on these mythical functions is now relegated to mathematics textbooks\cite{Greenhill,WW} and mathematical handbooks\cite{HMF_Jacobi,HMF_Weiers} with notations and conventions that are often contradictory or difficult to understand by physicists. The purpose of this paper is thus to (re)introduce the Jacobi and Weierstarss elliptic functions to a new generation of physics students through a series of standard problems found in classical mechanics.

Many problems in classical mechanics involve calculating of one of the following integrals.\cite{Taylor} The time integral
\begin{equation}
t(x) \;=\; \pm \int_{x_{0}}^{x}\; \frac{dy}{\sqrt{(2/m)\,[E - U(y)]}}
\label{eq:1D_problems}
\end{equation}
arises in solutions $x(t)$ of one-dimensional problems associated with motion of a particle of mass $m$ and constant total energy $E$ in a time-independent potential $U(x)$, where the initial condition $x_{0}$ may be chosen to correspond to a root of the turning-point equation $E = U(x)$. The orbit integral
\begin{equation}
\theta(s) \;=\; \pm \int_{s_{0}}^{s}\; \frac{d\sigma}{\sqrt{(2\mu/\ell^{2})[E - U(\sigma^{-1}) - \ell^{2}\sigma^{2}/2\mu]}}
\label{eq:central_problems} 
\end{equation}
arises in solutions $r(\theta) \equiv 1/s(\theta)$ of central-force problems involving the motion of a (fictitious) particle of (reduced) mass 
$\mu$, with constant total energy $E$ and angular momentum $\ell = \mu r^{2}\dot{\theta}$, in a central potential $U(r)$, where the initial condition $s(0) = s_{0}$ is a turning point. The time integral
\begin{equation}
t(\theta) \;=\; \pm \int_{\cos\theta_{0}}^{\cos\theta}\;\frac{du}{\sqrt{(2/I_{1})\,(1 - u^{2})\,[E - V(u)]}}
\label{eq:rigid_problems}
\end{equation}
arises in solutions $\theta(t)$ of rigid-body dynamics in the Lagrangian representation, where $I_{1}$ denotes one principal component of the inertia tensor for a symmetric top $(I_{1} = I_{2} \neq I_{3})$ and the effective potential $V(\cos\theta)$ contains terms associated with conserved angular momenta associated with the ignorable Eulerian angles $\psi$ and $\varphi$.

Exact analytical solutions for these integrals exist only for certain potentials, in which cases the inversions $t(x) \rightarrow x(t)$, $\theta(s) 
\rightarrow s(\theta)$, and $t(\theta) \rightarrow \theta(t)$ can be expressed in terms of known functions. For example, exact solutions of the time integral (\ref{eq:1D_problems}) exist in terms of trigonometric (or singly-periodic) functions when the potential $U(x)$ is a quadratic polynomial in $x$. Trigonometric solutions of the orbit integral (\ref{eq:central_problems}), on the other hand, exist for the Kepler problem $U(r) = -\,k/r$ and the isotropic harmonic oscillator $U(r) = k\,r^{2}/2$. The purpose of the present paper is to explore exact analytic solutions of the integrals (\ref{eq:1D_problems})-(\ref{eq:rigid_problems}) expressed in terms of doubly-periodic functions called elliptic functions.\cite{WW} For example, exact solutions of the orbit integral (\ref{eq:central_problems}) for the central potential $U(r) = k\,r^{n}$ exist in terms of elliptic functions \cite{Whittaker} for $n = \pm\,6, \pm\,4, 1$, and $-3$. The time integral (\ref{eq:rigid_problems}), on the other hand, has a solution in terms of elliptic functions for the problem of the heavy symmetric top (of mass $M$) with one fixed point (located at a distance $h$ from the center of mass), where $V(\cos\theta) = 
Mgh\,\cos\theta$.

\subsection{Doubly-periodic elliptic functions}

\begin{figure}
\epsfysize=2in
\epsfbox{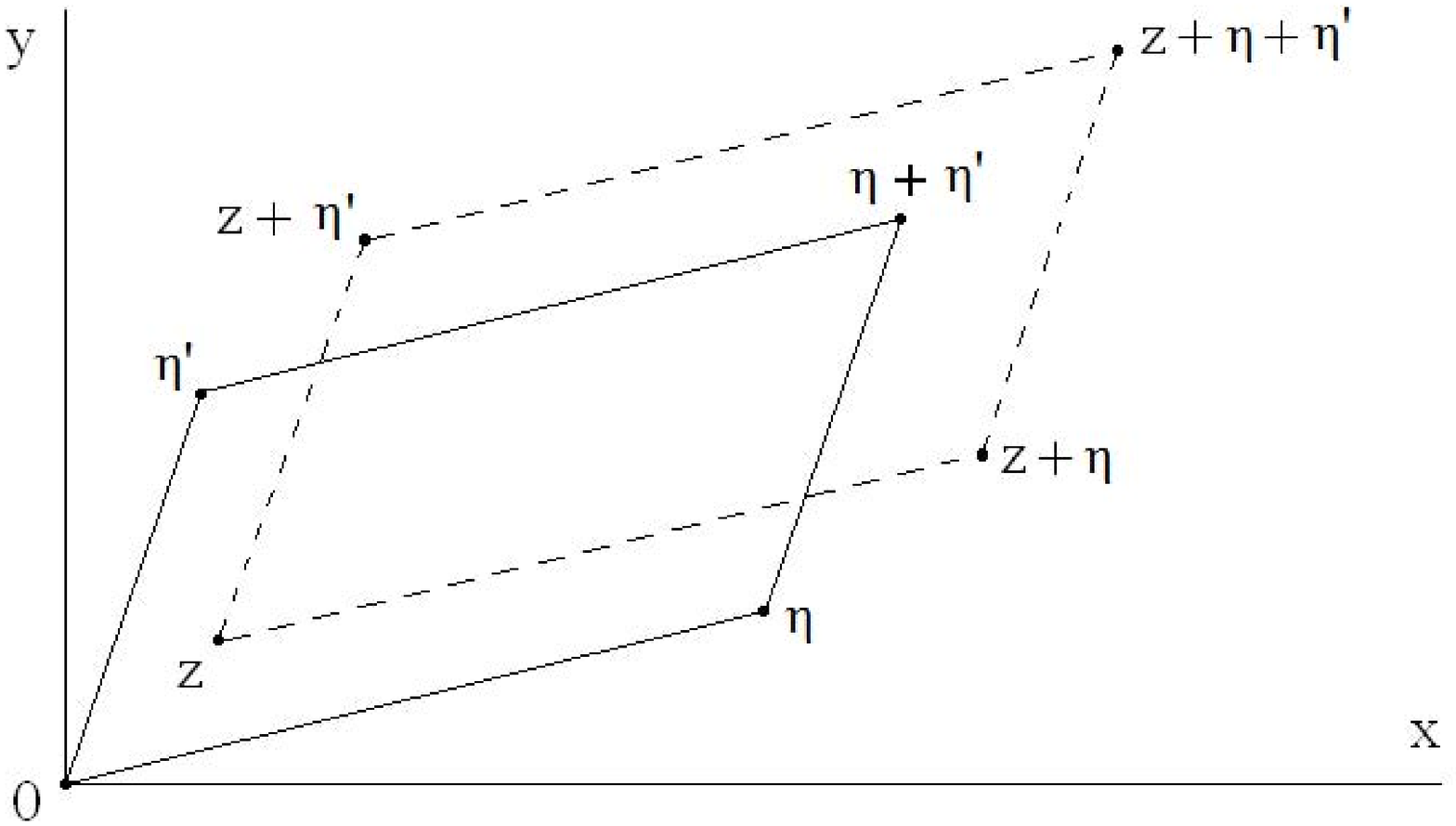}
\caption{Unit cell for a doubly-periodic elliptic function with periods $\eta$ and $\eta^{\prime}$. A doubly-periodic function $F(z)$ satisfies the property $F(z + m\,\eta + n\,\eta^{\prime}) \;=\; F(z)$ for $m, n = 0, \pm 1, \pm 2, ...$ (but not $m = 0 = n$).}
\label{fig:unit_cell}
\end{figure}

A function $F(z)$ is said to be doubly-periodic, with periods $\eta$ and $\eta^{\prime}$ (where the complex-valued ratio 
$\eta^{\prime}/\eta$ has a positive-definite imaginary part), if
\[ F(z + m\,\eta + n\,\eta^{\prime}) \;=\; F(z), \]
for $m, n = 0, \pm 1, \pm 2, ...$ (but not $m = 0 = n$). Figure \ref{fig:unit_cell} shows the {\it unit cell} (or fundamental period parallelogram) with corners at $z = 0, \eta, \eta^{\prime}$, and $\eta + \eta^{\prime}$ in the complex plane. We note that, in the limit $|\eta^{\prime}| \rightarrow 
\infty$, the function $F(z)$ becomes singly-periodic with period $\eta$. Elliptic functions are doubly-periodic functions with 2 simple zeroes per unit cell and either a second-order pole (Weierstrass elliptic function) or two first-order poles (Jacobi elliptic function). Note that there are no elliptic functions of first order and that there are no multiply-periodic functions with more than two periods. \cite{WW}

Elliptic functions $y(x; {\bf a})$ are defined as solutions of the nonlinear ordinary differential equation
\[ \left( \frac{dy}{dx}\right)^{2} \;=\; a_{4}\,y^{4} + a_{3}\,y^{3} + a_{2}\,y^{2} + a_{1}\,y + a_{0}, \]
where ${\bf a} \equiv (a_{0},a_{1}, ..., a_{4})$ are constant coefficients. This equation can be formally solved by finding the inverse function
\[ x(y;{\bf a}) \;=\; x_{0} \pm \int_{y_{0}({\bf a})}^{y}\frac{ds}{\sqrt{a_{4}\,s^{4} + a_{3}\,s^{3} + a_{2}\,s^{2} + a_{1}\,s + a_{0}}}, \]
where $y_{0}({\bf a})$ is a root of the quartic polynomial $a_{4}\,y^{4} + a_{3}\,y^{3} + a_{2}\,y^{2} + a_{1}\,y + a_{0}$ and $x(y_{0};{\bf a}) = 
x_{0}$.  Jacobi elliptic functions are defined in terms of the quartic polynomial $(1 - y^{2})\,(1 - m\,y^{2})$, where $m$ is a positive constant, while Weierstrass elliptic functions are defined in terms of the cubic polynomial $4\,y^{3} - g_{2}\,y - g_{3}$, where $g_{2}$ and $g_{3}$ are constants. There is a connection between the Jacobi and Weierstrass elliptic functions \cite{WW} that will be exploited later in Sec.~\ref{subsec:plan_w} (see also Appendices \ref{sec:W_J} and \ref{sec:math}).

\subsection{Organization}

The remainder of the paper is organized as follows. In Sec.~\ref{sec:Jacobi_elliptic}, we present the Jacobi elliptic functions and discuss the Seiffert spherical spiral\cite{WW,Erdos_Seiffert} as a mathematical introduction of their doubly-periodic nature. Next, we discuss exact solutions to the physical problems of (i) the periodic motion in a quartic potential, (ii) the planar pendulum, and (iii) Euler's equations for a force-free asymmetric top. In Sec.~\ref{sec:Weierstrass_elliptic}, we present the Weierstrass elliptic functions and discuss exact solutions to the physical problems of (i) the motion in a cubic potential, (ii) the planar pendulum (which demonstrates the connection between the Jacobi and Weierstrass elliptic functions), (iii) the spherical pendulum, and (iv) the motion of a heavy symmetric top with one fixed point. In Sec.~\ref{sec:KdV}, we discuss one interesting application of elliptic functions in terms of the travelling-wave solutions of nonlinear partial differential equations. We summarize our work in Sec.~\ref{sec:summary} and present mathematical details in Appendices \ref{sec:W_J} and \ref{sec:math}. Lastly, we note that the notation used for the Jacobi and Weierstrass elliptic functions in our paper is partly based on the standard material presented elsewhere.\cite{HMF_Jacobi,HMF_Weiers}

\section{\label{sec:Jacobi_elliptic}Jacobi Elliptic Functions}

We begin our introduction of elliptic functions with the more familiar Jacobi elliptic functions. The Jacobi elliptic function ${\rm sn}(z\,|\,m)$ is defined in terms of the inverse-function formula
\begin{eqnarray} 
z & = & \int_{0}^{\varphi}\,\frac{d\theta}{\sqrt{1 \;-\; m\,\sin^{2}\theta}} \nonumber \\
 & = & \int_{0}^{\sin\varphi}\,\frac{dy}{\sqrt{(1 - y^{2})\,(1 - m\,y^{2})}} \nonumber \\
 & \equiv & {\rm sn}^{-1}(\sin\varphi\,|\, m),
\label{eq:sn_def}
\end{eqnarray}
where the modulus $m$ is a positive number and the amplitude $\varphi$ varies from $0$ to $2\pi$. From this definition, we easily check that ${\rm sn}^{-1}(\sin\varphi\,|\, 0) = \sin^{-1}(\sin\varphi) = \varphi$. The solution to the differential equation 
\begin{equation}
\left( \frac{dy}{dz}\right)^{2} \;=\; \left( 1 - y^{2} \right)\;\left( 1 - m\,y^{2} \right) 
\label{eq:Jy_ode}
\end{equation}
is expressed in terms of the Jacobi elliptic function
\begin{equation}
y(z) \;=\; \left\{ \begin{array}{lr}
{\rm sn}(z|m) & ({\rm for}\;\; m \;<\; 1), \\
 & \\
m^{-1/2}\;{\rm sn}\left(m^{1/2}\,z\;|\;m^{-1}\right) & \;\;\;({\rm for}\;\; m \;>\; 1).
\end{array} \right.
\label{eq:y_Jsol}
\end{equation}
By using the transformation $y = \sin\varphi$, the Jacobi differential equation (\ref{eq:Jy_ode}) becomes
\begin{equation}
\left( \frac{d\varphi}{dz}\right)^{2} \;=\; 1 - m\,\sin^{2}\varphi,
\label{eq:Jphi_ode}
\end{equation}
and the solution to this equation is $\varphi(z) = \sin^{-1}[ {\rm sn}(z|m)]$ for $m < 1$.

\begin{figure}
\epsfysize=2in
\epsfbox{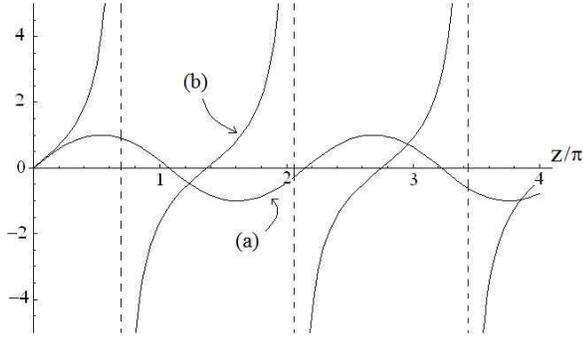}
\caption{Plots of (a) ${\rm sn}(z|m)$ and (b) $-i\,{\rm sn}(iz|m)$ for $m = 1/16$ showing the real and imaginary periods 
$4\,K(m)$ and $4\,i\,K^{\prime}(m)$.}
\label{fig:Jacobi}
\end{figure}

\begin{figure}
\epsfysize=2in
\epsfbox{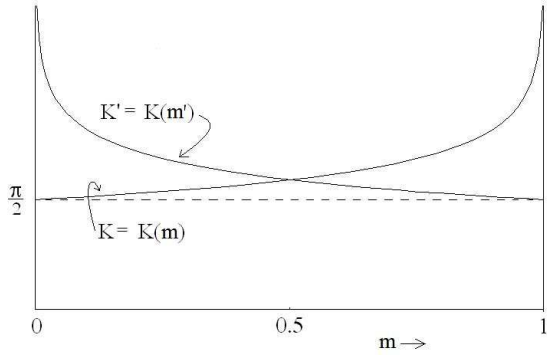}
\caption{Plots of the quarter periods $K = K(m)$ and $K^{\prime} = K(m^{\prime}) = K(1 - m)$.}
\label{fig:Jacobi_per}
\end{figure}

The function ${\rm sn}(z|m)$ has a purely-real period $4\,K$, where the quarter-period $K$ is defined as
\begin{equation} 
K \;\equiv\; K(m) \;=\; \int_{0}^{\pi/2}\,\frac{d\theta}{\sqrt{1 \;-\; m\,\sin^{2}\theta}}
\label{eq:K_def}
\end{equation}
and a purely-imaginary period $4\,iK'$, where the quarter-period $K'$ is defined as (with the complementary modulus $m^{\prime} \equiv 1 - m$)
\begin{equation}
i\,K'\;\equiv\; i\,K(m') \;=\; i\;\int_{0}^{\pi/2}\,\frac{d\theta}{\sqrt{1 \;-\; m'\,\sin^{2}\theta}}.
\label{eq:Kprime_def}
\end{equation}
Figure \ref{fig:Jacobi} shows plots of ${\rm sn}\,z$ and $-i\,{\rm sn}(iz)$ for $m = 1/16$, which exhibit both a real period and an imaginary period. Note that, while the Jacobi elliptic function ${\rm sn}\,z$ alternates between $-1$ and 
$+1$ for real values of $z$ (with zeroes at $2n\,K$), it also exhibits singularities for imaginary values of $z$ at $(2n + 1)\,iK'$ ($n = 0, 1, ...$). Furthermore, as $m \rightarrow 0$ (and $m' \rightarrow 1$), we find $K \rightarrow \pi/2$ (or $4\,K \rightarrow 2\pi$) and $|K'| \rightarrow \infty$ (see Figure \ref{fig:Jacobi_per}), and so ${\rm sn}\,z \rightarrow \sin z$ becomes singly-periodic.

\begin{figure}
\epsfysize=2in
\epsfbox{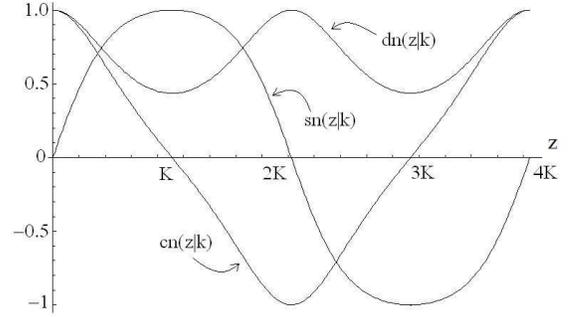}
\caption{Plots of ${\rm sn}(z|m)$, ${\rm cn}(z|m)$, and ${\rm dn}(z|m)$ from $z = 0$ to $4\,K(m)$ for $m = (0.9)^{2}$.}
\label{fig:Jacobi_scd}
\end{figure}

The additional Jacobi elliptic functions ${\rm cn}(z\,|\,m)$ and ${\rm dn}(z\,|\,m)$ are defined as
\begin{eqnarray}
z & = & \int_{{\rm cn}(z|m)}^{1}\,\frac{dy}{\sqrt{(1 - y^{2})\,(m' + m\,y^{2})}}, \label{eq:cn_def} \\
  & = & \int_{{\rm dn}(z|m)}^{1}\,\frac{dy}{\sqrt{(1 - y^{2})\,(y^{2} - m')}}, \label{eq:dn_def}
\end{eqnarray}
with the properties ${\rm cn}\,z \equiv {\rm cn}(z|m) = \cos\varphi$, ${\rm dn}\,z \equiv {\rm dn}(z|m) = 
\sqrt{1 - m\,\sin^{2}\varphi}$, and ${\rm sn}^{2}z + {\rm cn}^{2}z = 1 = {\rm dn}^{2}z + m\,{\rm sn}^{2}z$. The Jacobi elliptic functions ${\rm cn}\,z$ and ${\rm dn}\,z$ are also doubly-periodic with periods $4\,K$ and $4i\,K'$ (see Figure \ref{fig:Jacobi_scd}). 

The following properties are useful. First, we find the limits:
\begin{equation}
\left( \begin{array}{c}
{\rm sn}(z|0) \\
{\rm cn}(z|0) \\
{\rm dn}(z|0)
\end{array} \right) \;=\; \left( \begin{array}{c}
\sin z \\
\cos z \\
1
\end{array} \right) 
\label{eq:scd_0}
\end{equation}
and
\begin{equation} 
\left( \begin{array}{c}
{\rm sn}(z|1) \\
{\rm cn}(z|1) \\
{\rm dn}(z|1)
\end{array} \right) \;=\; \left( \begin{array}{c}
\tanh z \\
{\rm sech}\,z \\
{\rm sech}\,z
\end{array} \right).
\label{eq:scd_1}
\end{equation}
Next, we find the derivatives with respect to the argument $z$:
\begin{equation}
\left. \begin{array}{rcl}
{\rm sn}^{\prime}(z|m) & = & {\rm cn}(z|m)\; {\rm dn}(z|m) \\
{\rm cn}^{\prime}(z|m) & = & -\;{\rm sn}(z|m)\; {\rm dn}(z|m) \\
{\rm dn}^{\prime}(z|m) & = & -\;m\;{\rm cn}(z|m)\; {\rm sn}(z|m)
\end{array} \right\},
\label{eq:scd_der}
\end{equation}
and, if $m > 1$, the identities:
\begin{equation}
\left. \begin{array}{rcl}
{\rm sn}(z|m) & = & m^{-1/2}\; {\rm sn}(m^{1/2}\,z|m^{-1}) \\
{\rm cn}(z|m) & = & {\rm dn}(m^{1/2}\,z|m^{-1}) \\
{\rm dn}(z|m) & = & {\rm cn}(m^{1/2}\,z|m^{-1})
\end{array} \right\}.
\label{eq:scd_big}
\end{equation}
We now turn our attention to solving mathematical and physical problems using the Jacobi elliptic functions (\ref{eq:sn_def}) and 
(\ref{eq:cn_def})-(\ref{eq:dn_def}).

\subsection{Seiffert spherical spiral}

\begin{figure}
\epsfysize=2in
\epsfbox{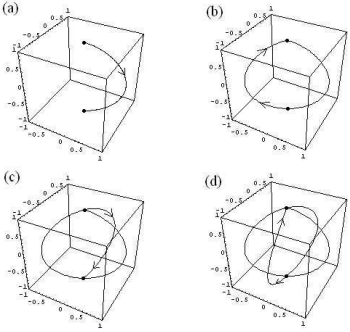}
\caption{Seiffert Spherical Spiral: Plot of the vector (\ref{eq:seiffert_unit}) on the surface of the unit sphere for $k = 0.15$ from $s = 0$ to (a) $s = 2K$, (b) $s = 4K$, (c) $s = 6K$, and (d) $s = 8K$.}
\label{fig:Seiffert}
\end{figure}

\begin{figure}
\epsfysize=2in
\epsfbox{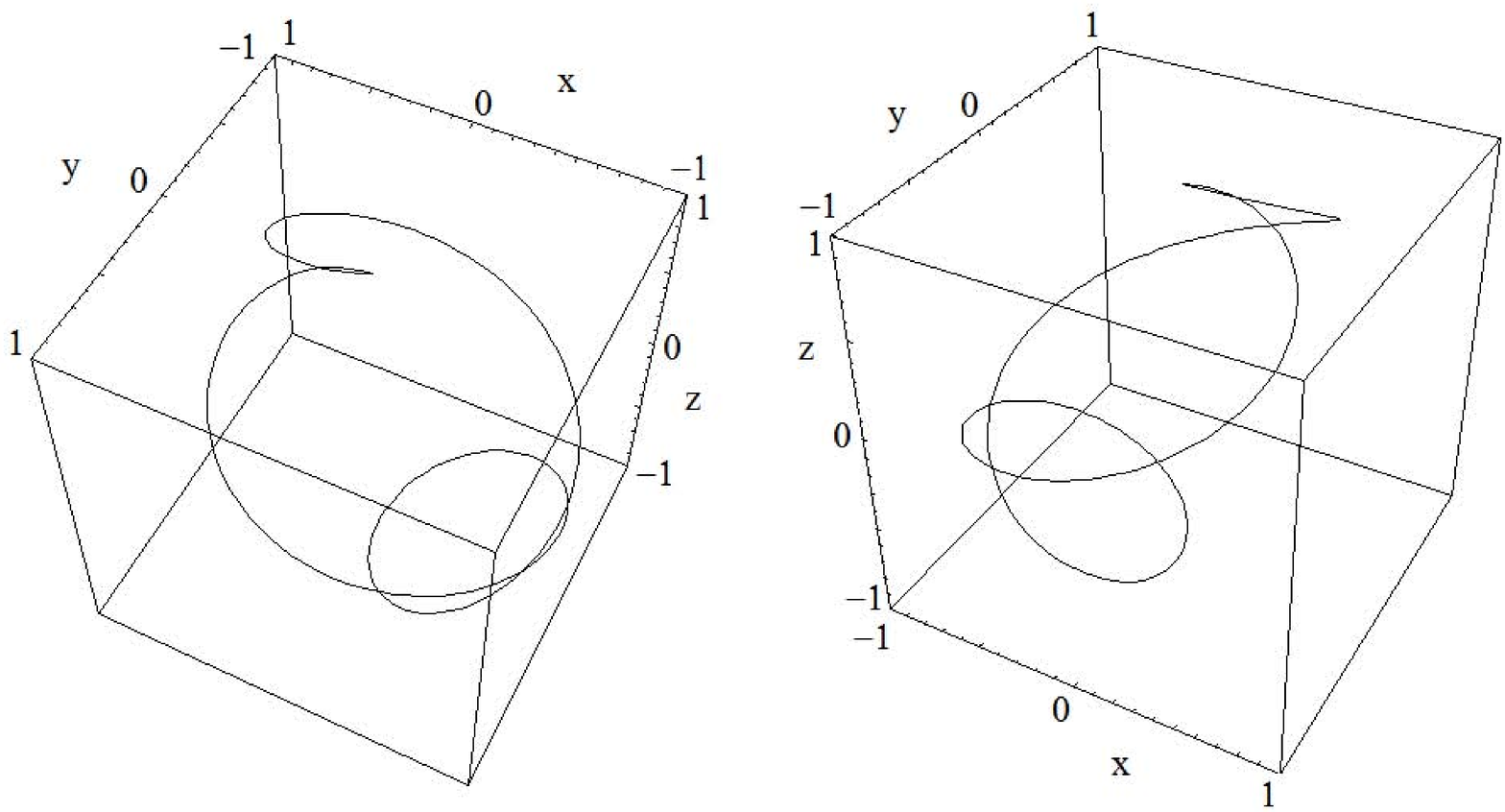}
\caption{Two different views of the Seiffert Spherical Spiral for $k = 0.95$ from $s = 0$ to $s = 4K$.}
\label{fig:Seiffert_2}
\end{figure}

A simple example that clearly demonstrates the periodicity of the Jacobi elliptic functions ${\rm sn}\,z$ and ${\rm cn}\,z$ is given by the {\it Seiffert spherical spiral},\cite{WW,Erdos_Seiffert} defined as a periodic curve on the unit sphere and constructed as follows. First, we use the cylindrical metric $ds^{2} = d\rho^{2} + \rho^{2}\,d\varphi^{2} + dz^{2}$, with $z = \sqrt{1 - \rho^{2}}$ and the azimuthal angle $\varphi(s) \equiv 
k\,s$ is parametrized by the arc length $s$ (assuming that the initial point of the curve is $\rho = 0$, $\varphi = 0$, and $z = 1$). Hence, we readily find $ds^{2} = [(1 - \rho^{2})\,(1 - k^{2}\,\rho^{2})]^{-1}d\rho^{2}$, which leads to (with $0 < m \equiv k^{2} < 1$)
\[ s \;=\; \int_{0}^{\rho}\;\frac{dy}{\sqrt{(1 - y^{2})\,(1 - m\,y^{2})}} \;\equiv\; {\rm sn}^{-1}(\rho|m), \]
and, thus, we obtain the Jacobi elliptic solutions
\begin{equation} 
\left. \begin{array}{rcl}
\rho(s) & = & {\rm sn}(s|m) \\
 &  & \\ 
z(s) & = & \sqrt{1 - \rho^{2}(s)} \;=\; {\rm cn}(s|m)
\end{array} \right\}.
\label{eq:Seiffert_sol}
\end{equation}
The case $m > 1$ is handled with the identities (\ref{eq:scd_big}). The Seiffert spherical spiral is generated by plotting on the unit sphere the path of the unit vector 
\begin{equation}
\wh{{\sf r}}(s) \;=\; {\rm sn}(s|k^{2}) \left[\; \cos(ks)\;\wh{{\sf x}} \;+\; \sin(ks)\;\wh{{\sf y}} \;\right] \;+\; 
{\rm cn}(s|k^{2})\;\wh{{\sf z}} 
\label{eq:seiffert_unit}
\end{equation}
as a function of $s$ (see Figure \ref{fig:Seiffert}). Note that at each value $4n\,K$ $(n = 1, 2,...)$, the orbit returns to the initial point at $\rho = 0$ and $z = 1$. Figure \ref{fig:Seiffert_2} shows the complex periodic nature of the Seiffert-spiral orbit for $k = 0.95$.

\subsection{Motion in a quartic potential}

\begin{figure}
\epsfysize=2in
\epsfbox{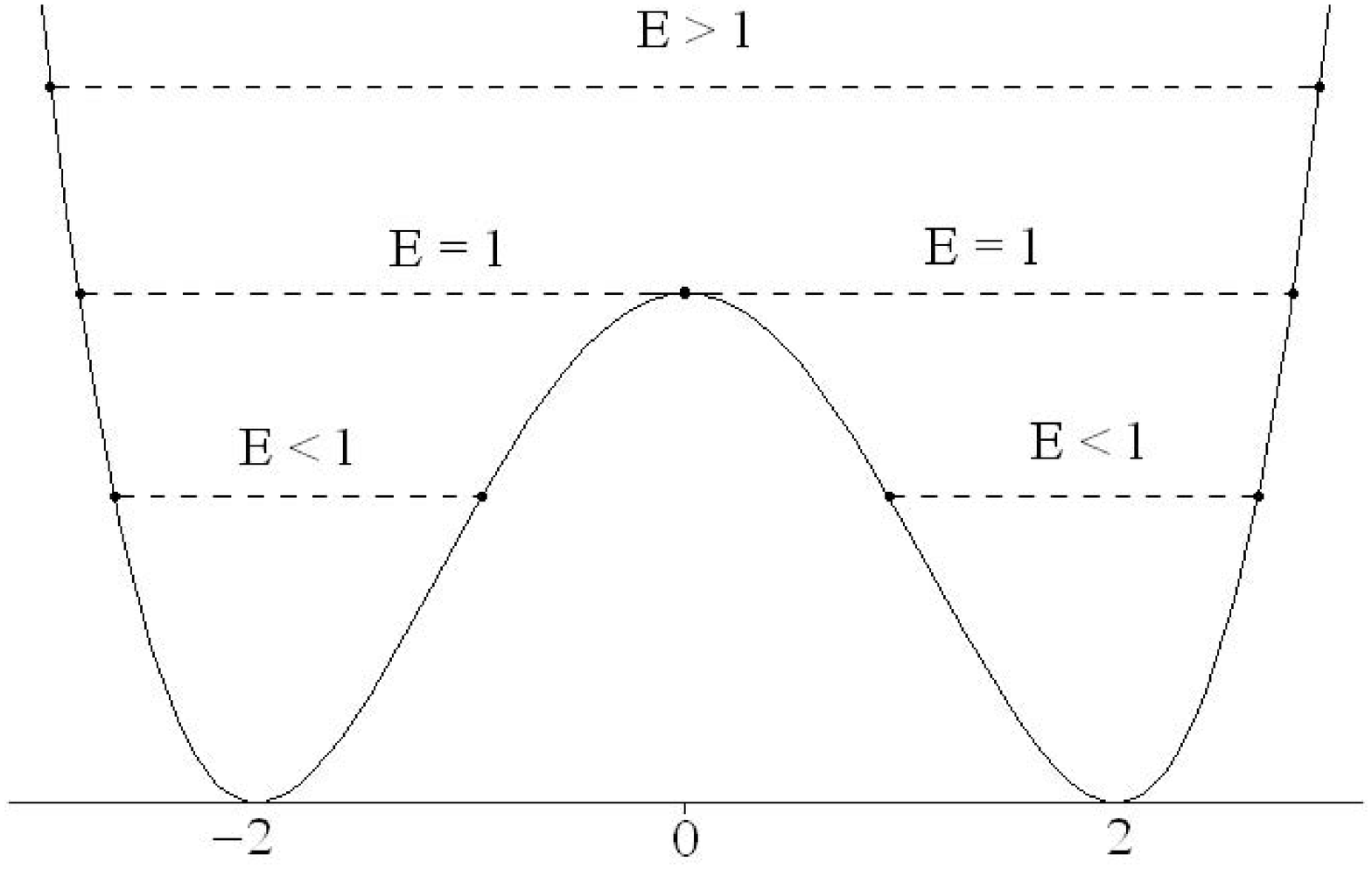}
\caption{Quartic potential $U(x) = 1 - x^{2}/2 + x^{4}/16$ showing orbits with $E > 1$, $E = 1$ (separatrix) and $E < 1$.}
\label{fig:quartic_pot}
\end{figure}

\begin{figure}
\epsfysize=2in
\epsfbox{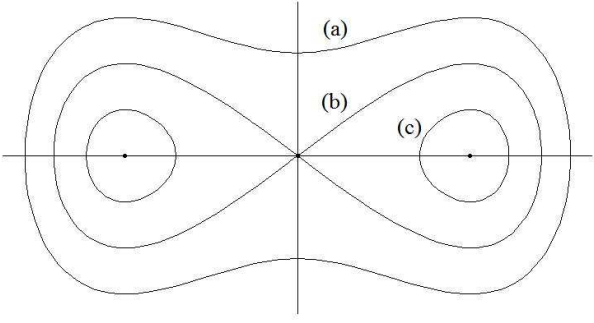}
\caption{Phase portait for orbits (\ref{eq:quartic_large})-(\ref{eq:quartic_small}) of the quartic potential $U(x) = 1 - x^{2}/2 + x^{4}/16$ for (a) ${\sf e} > 1$, (b) ${\sf e} = 1$ (separatrix), and (c) ${\sf e} < 1$.}
\label{fig:quartic_orb}
\end{figure}

Our next example looks at particle orbits in the (dimensionless) quartic potential $U(x) = 1 - x^{2}/2 + x^{4}/16$ shown in Figure \ref{fig:quartic_pot}. Here, the turning points for $E \equiv {\sf e}^{2} = U(x)$ are 
\begin{equation}
\left. \begin{array}{lr}
\pm\,2\;\sqrt{1 + {\sf e}} & ({\rm for}\;{\sf e} > 1) \\
 & \\
0 \;{\rm and}\; \pm\,\sqrt{8} & ({\rm for}\;{\sf e} = 1) \\
 & \\
\pm\,2\;\sqrt{1 \pm {\sf e}} & ({\rm for}\;{\sf e} < 1)
\end{array} \right\}.
\label{eq:tp_quartic}
\end{equation}
Each orbit is solved in terms of the integral (\ref{eq:1D_problems}) using the initial condition $x_{0} = 2\,
\sqrt{1 + {\sf e}}$ with the initial velocity $\dot{x}_{0} < 0$:
\begin{eqnarray}
t(x) & = & -\;\int_{2\,\sqrt{1 + {\sf e}}}^{x}\;\frac{dy}{\sqrt{2\,({\sf e}^{2} - 1) \;+\; y^{2}\,(1 - y^{2}/8)}} \nonumber \\
 & = & -\;\int_{2\,\sqrt{1 + {\sf e}}}^{x}\;\frac{\sqrt{8}\;dy}{\sqrt{[4\,({\sf e} + 1) - y^{2}]\,[y^{2} + 4\,({\sf e} - 1)]}} \nonumber \\
 & = & \frac{1}{\sqrt{{\sf e}}}\;\int_{0}^{\Phi(x)}\; \frac{d\varphi}{\sqrt{1 \;-\; m\;\sin^{2}\varphi}},
\label{eq:quartic_first}
\end{eqnarray}
where $m \equiv (1 + {\sf e})/2{\sf e}$ while we used the trigonometric substitution $y = 2\,\sqrt{1 + {\sf e}}\,\cos\varphi$ with
\begin{equation}
\Phi(x) \;\equiv\; \cos^{-1}\left[\frac{x}{2\sqrt{1 + {\sf e}}}\right] 
\label{eq:Phi4_def}
\end{equation}
to obtain the last expression in Eq.~(\ref{eq:quartic_first}). The Jacobi elliptic solutions obtained from Eq.~(\ref{eq:quartic_first}) are shown in Figure \ref{fig:quartic_orb} for the orbit (a), with ${\sf e} > 1$, the separatrix orbit (b), with ${\sf e} = 1$, and the orbit (c), with ${\sf e} < 1$.

For ${\sf e} > 1$ (i.e., $m < 1$), we use Eq.~(\ref{eq:sn_def}) to find 
\[ \sin\Phi(x) \;=\; {\rm sn}(\sqrt{{\sf e}}\,t|m) \;=\; \sqrt{1 \;-\; \frac{x^{2}(t)}{4\,(1 + {\sf e})}}, \]
which yields the phase-portrait coordinates $(x,\dot{x})$:
\begin{equation}
\left. \begin{array}{rcl}
x(t) & = & 2\,\sqrt{1 + {\sf e}}\;\;{\rm cn}(\sqrt{{\sf e}}\,t|m) \\
 &  & \\
\dot{x}(t) & = & -\;2\,\sqrt{{\sf e}\,(1 + {\sf e})}\;\;{\rm sn}(\sqrt{{\sf e}}\,t|m)\,{\rm dn}(\sqrt{{\sf e}}\,t|m)
\end{array} \right\},
\label{eq:quartic_large}
\end{equation}
where the velocity $\dot{x}(t)$ is obtained by using Eq.~(\ref{eq:scd_der}). For ${\sf e} = 1$ (i.e., the separatrix orbit with $m = 1$), the phase-portrait coordinates become
\begin{equation}
\left. \begin{array}{rcl}
x(t) & = & \sqrt{8}\;\;{\rm sech}\,t \\
 &  & \\
\dot{x}(t) & = & -\;\sqrt{8}\;\;{\rm sech}\,t\;\tanh t
\end{array} \right\},
\label{eq:quartic_sep}
\end{equation}
where the limits (\ref{eq:scd_1}) were applied to Eq.~(\ref{eq:quartic_large}). Lastly, for ${\sf e} < 1$ (i.e., $m > 1$), we apply the relations 
(\ref{eq:scd_big}) on Eq.~(\ref{eq:quartic_large}) to obtain
\begin{equation}
\left. \begin{array}{rcl}
x(t) & = & 2\,\sqrt{1 + {\sf e}}\;\;{\rm dn}(\tau\,|m^{-1}) \\
 &  & \\
\dot{x}(t) & = & -\;\sqrt{8}\;{\sf e}\;\;{\rm sn}(\tau\,|m^{-1})\,{\rm cn}(\tau\,|m^{-1})
\end{array} \right\},
\label{eq:quartic_small}
\end{equation}
where $\tau = t\;\sqrt{(1 + {\sf e})/2}$. The orbits (\ref{eq:quartic_large})-(\ref{eq:quartic_small}) are combined to yield the phase portrait for the quartic potential shown in Figure \ref{fig:quartic_orb}.

\subsection{\label{subsec:ppend}Planar pendulum}

\begin{figure}
\epsfysize=2in
\epsfbox{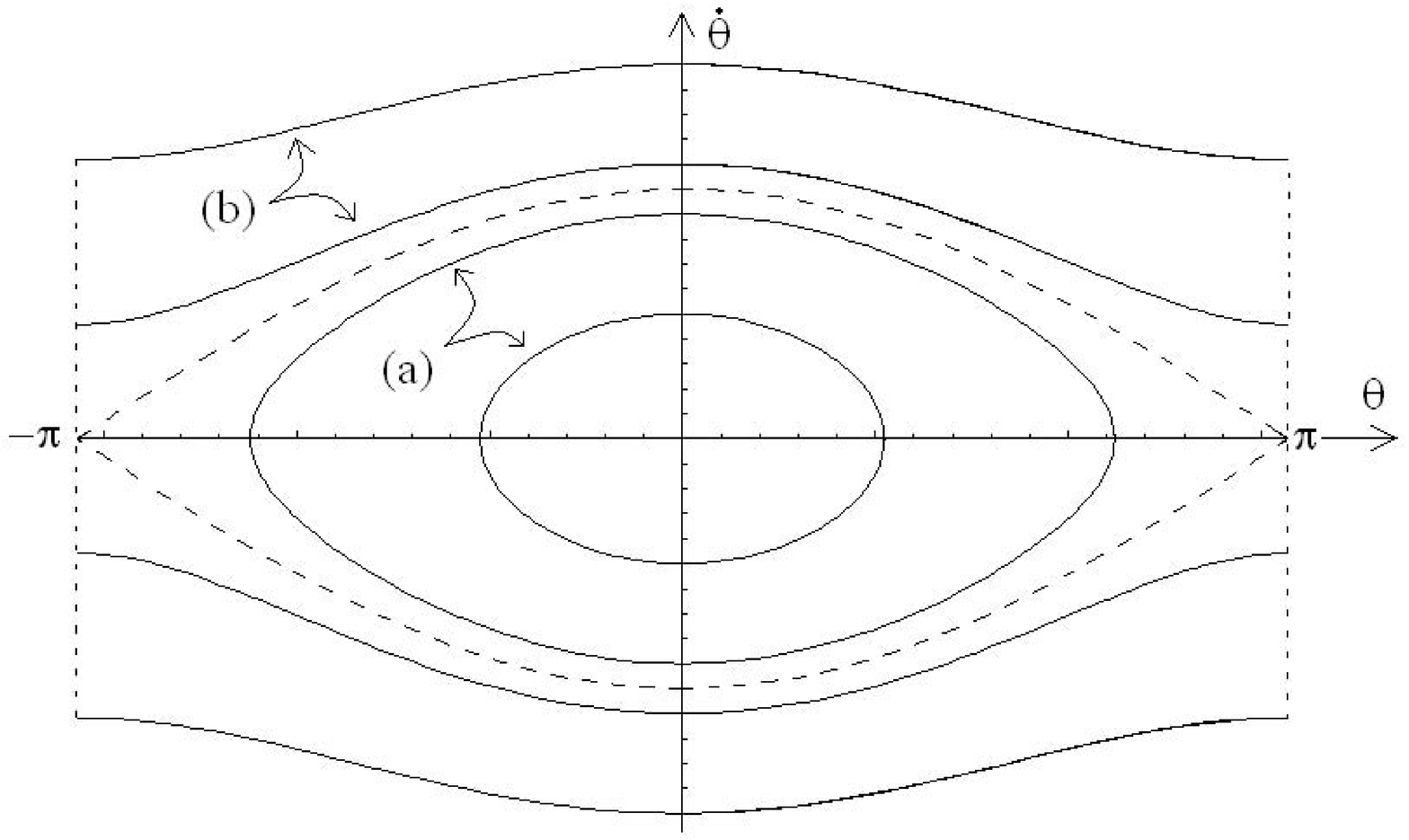}
\caption{Phase portait for pendulum orbits: (a) bounded (oscillation) orbits ($m < 1$) and (b) unbounded (libration) orbits ($m > 1$). The separatrix orbit (dashed line) separates oscillations from librations (which are periodic since 
$-\pi$ is identical to $\pi$).}
\label{fig:pendulum_orb}
\end{figure}

As a first physical example, we consider the well-known planar pendulum with a normalized energy $\epsilon\,\omega_{0}^{2}$ $(\omega_{0}^{2} = g/L)$ expressed as
\begin{eqnarray}
\epsilon\,\omega_{0}^{2} & = & \frac{1}{2}\,\dot{\theta}^{2} \;+\; \omega_{0}^{2}\,(1 - \cos\theta) \nonumber \\
 & = & 2\,\omega_{0}^{2}\;\left( \varphi'^{2} \;+\; \sin^{2}\varphi\right),
\label{eq:ppendulum_eq}
\end{eqnarray}
where $\varphi \equiv \theta/2$ and $\varphi'(\tau) \equiv \omega_{0}^{-1}\dot{\varphi}$. The normalized time (assuming that $\varphi = 0$ at 
$\tau = 0$) is thus expressed as
\begin{equation} 
\tau(\varphi) \;\equiv\; \omega_{0}\,t(\varphi) \;=\; \int_{0}^{\varphi}\; \frac{d\phi}{\sqrt{(\epsilon/2) - \sin^{2}\phi}}.
\label{eq:tau_pend}
\end{equation}
From Eq.~(\ref{eq:ppendulum_eq}), the dimensionless energy $\epsilon$ can either be (I) $0 < \epsilon < 2$ (i.e., $\varphi^{\prime}$ vanishes along the orbit) or (II) $\epsilon > 2$ (i.e., $\varphi^{\prime}$ does not vanish along the orbit). In case (I), we set $\epsilon \equiv 2\,\sin^{2}\alpha$ and $\sin\phi = \sin\alpha\,\sin\chi$ in Eq.~(\ref{eq:tau_pend}) to obtain \cite{Landau} 
\begin{eqnarray*} 
\tau(\varphi) & = & \int_{0}^{\arcsin(m^{-1/2}\,\sin\varphi)}\;\frac{d\chi}{\sqrt{1 - m\,\sin^{2}\chi}} \\
 & \equiv & {\rm sn}^{-1}\left(m^{-1/2}\sin\varphi\,|\,m \right),
\end{eqnarray*}
where the modulus is $m \equiv \sin^{2}\alpha = \epsilon/2$. Hence, the solution for case (I) is expressed in terms of the Jacobi elliptic function
\begin{equation} 
\varphi(\tau) \;=\; \sin^{-1}\left[\;m^{1/2}\;{\rm sn}(\tau\,|\,m)\;\right].
\label{eq:pen_lib}
\end{equation}
In the limit $m \ll 1$ (i.e., $\epsilon \ll 2$), we use the limit (\ref{eq:scd_0}) to obtain the simple harmonic solution $\theta(\tau) = 
\sqrt{2\,\epsilon}\,\sin\tau$. 

In case (II), we have $m^{-1} = 2/\epsilon < 1$ and obtain
\begin{eqnarray*}
\tau(\varphi) & = & \sqrt{\frac{2}{\epsilon}}\;\int_{0}^{\varphi}\;\frac{d\phi}{\sqrt{1 - m\,\sin^{2}\phi}} \\
 & \equiv & m^{-1/2}\;{\rm sn}^{-1}(\sin\varphi\,|\,m^{-1}).
\end{eqnarray*}
Hence, the solution for case (II) is expressed in terms of the Jacobi elliptic function
\begin{equation}
\varphi(\tau) \;=\; \sin^{-1}\left[{\rm sn}(m^{1/2}\tau\,|\,m^{-1})\;\right],
\label{eq:pen_rot}
\end{equation}
which follows from property (\ref{eq:scd_big}). In the limit $m \rightarrow 1$, both solutions (\ref{eq:pen_lib}) and (\ref{eq:pen_rot}) coincide with the {\it separatrix} solution
\begin{equation}
\sin\varphi(\tau) \;=\; \tanh \tau, 
\label{eq:pen_sep}
\end{equation}
which is expressed in terms of singly-periodic (hyperbolic) trigonometric functions (with imaginary period). Since 
$\varphi \rightarrow \pm \pi/2$ ($\theta \rightarrow \pm\,\pi$) as $\tau \rightarrow \pm \infty$, the period of the pendulum on the separatrix orbit is infinite. The pendulum orbits (\ref{eq:pen_lib})-(\ref{eq:pen_sep}) are shown in Figure \ref{fig:pendulum_orb}.

We will return to this example in Sec.~\ref{subsec:plan_w} where we solve the problem of the planar pendulum in terms of the Weierstrass elliptic function and establish a connection between the Jacobi and Weierstrass elliptic functions.

\subsection{Force-free asymmetric top}

As a second physical example, we consider the Euler equations for a force-free asymmetric top (with principal moments of inertia $I_{1} > I_{2} > 
I_{3}$):\cite{Landau}
\begin{equation}
\left. \begin{array}{rcl}
I_{1}\,\dot{\omega}_{1} & = & (I_{2} - I_{3})\;\omega_{2}\,\omega_{3} \\
I_{2}\,\dot{\omega}_{2} & = & -\;(I_{1} - I_{3})\;\omega_{1}\,\omega_{3} \\
I_{3}\,\dot{\omega}_{3} & = & (I_{1} - I_{2})\;\omega_{1}\,\omega_{2}
\end{array} \right\},
\label{eq:euler_asym}
\end{equation}
where the angular velocity $\vb{\omega} = \omega_{1}\,\wh{{\sf 1}} + \omega_{2}\,\wh{{\sf 2}} + \omega_{3}\,\wh{{\sf 3}}$ is decomposed in terms of its components along the principal axes of inertia. The conservation laws of kinetic energy
\begin{equation}
\kappa \;=\; \frac{1}{2}\; \left( I_{1}\,\omega_{1}^{2} \;+\; I_{2}\,\omega_{2}^{2} \;+\; I_{3}\,\omega_{3}^{2} \right) \;\equiv\; \frac{1}{2}\;
I_{0}\,\Omega_{0}^{2},
\label{eq:K_asym} 
\end{equation}
and (squared) angular momentum 
\begin{equation}
\ell^{2} \;=\; I_{1}^{2}\,\omega_{1}^{2} \;+\; I_{2}^{2}\,\omega_{2}^{2} \;+\; I_{3}^{2}\,\omega_{3}^{2} \;\equiv\; I_{0}^{2}\,\Omega_{0}^{2},
\label{eq:L2_asym} 
\end{equation}
are expressed in terms of the parameters $I_{0} \equiv \ell^{2}/(2\,\kappa)$ and $\Omega_{0} \equiv 2\,\kappa/\ell$. These conservation laws can be used to introduce the following definitions
\begin{eqnarray}
\omega_{1}(\tau) & = & -\;\sqrt{\frac{I_{0}\,(I_{0} - I_{3})}{I_{1}\,(I_{1} - I_{3})}}\; \Omega_{0}\;\sqrt{1 - y^{2}(\tau)} \nonumber \\
 & \equiv & -\;\Omega_{1}(I_{0})\; \sqrt{1 - y^{2}(\tau)}, \label{eq:omega1_nu} \\
\omega_{2}(\tau) & = & \sqrt{\frac{I_{0}\,(I_{0} - I_{3})}{I_{2}\,(I_{2} - I_{3})}}\;\Omega_{0}\;y(\tau) \nonumber \\
 & \equiv & \Omega_{2}(I_{0})\;y(\tau), \label{eq:omega2_nu} \\
\omega_{3}(\tau) & = & \sqrt{\frac{I_{0}\,(I_{1} - I_{0})}{I_{3}\,(I_{1} - I_{3})}}\; \Omega_{0}\,\sqrt{1 - m\,y^{2}(\tau)} \nonumber \\
 & \equiv & \Omega_{3}(I_{0})\;\sqrt{1 - m\,y^{2}(\tau)}, \label{eq:omega3_nu}
\end{eqnarray}
where $\tau = [(I_{1} - I_{3})\,\Omega_{1}\,\Omega_{3}/(I_{2}\,\Omega_{2})]\,t$ is the dimensionless time and the modulus $m$ is defined as
\begin{equation}
m(I_{0}) \;\equiv\; \frac{(I_{0} - I_{3})\;(I_{1} - I_{2})}{(I_{2} - I_{3})\;(I_{1} - I_{0})}.
\label{eq:m_asym}
\end{equation}
By requiring that the modulus $m$ be positive, the parameter $I_{0}$ introduced in Eqs.~(\ref{eq:K_asym})-(\ref{eq:L2_asym}) must satisfy $I_{3} < I_{0} < I_{1}$ and, hence, $0 \leq m(I_{0}) \leq 1$ for $I_{3} \leq I_{0} \leq I_{2}$ and $m(I_{0}) > 1$ for $I_{2} < I_{0} < I_{1}$ (with $m \rightarrow \infty$ as $I_{0} \rightarrow I_{1}$).

\begin{figure}
\epsfysize=2in
\epsfbox{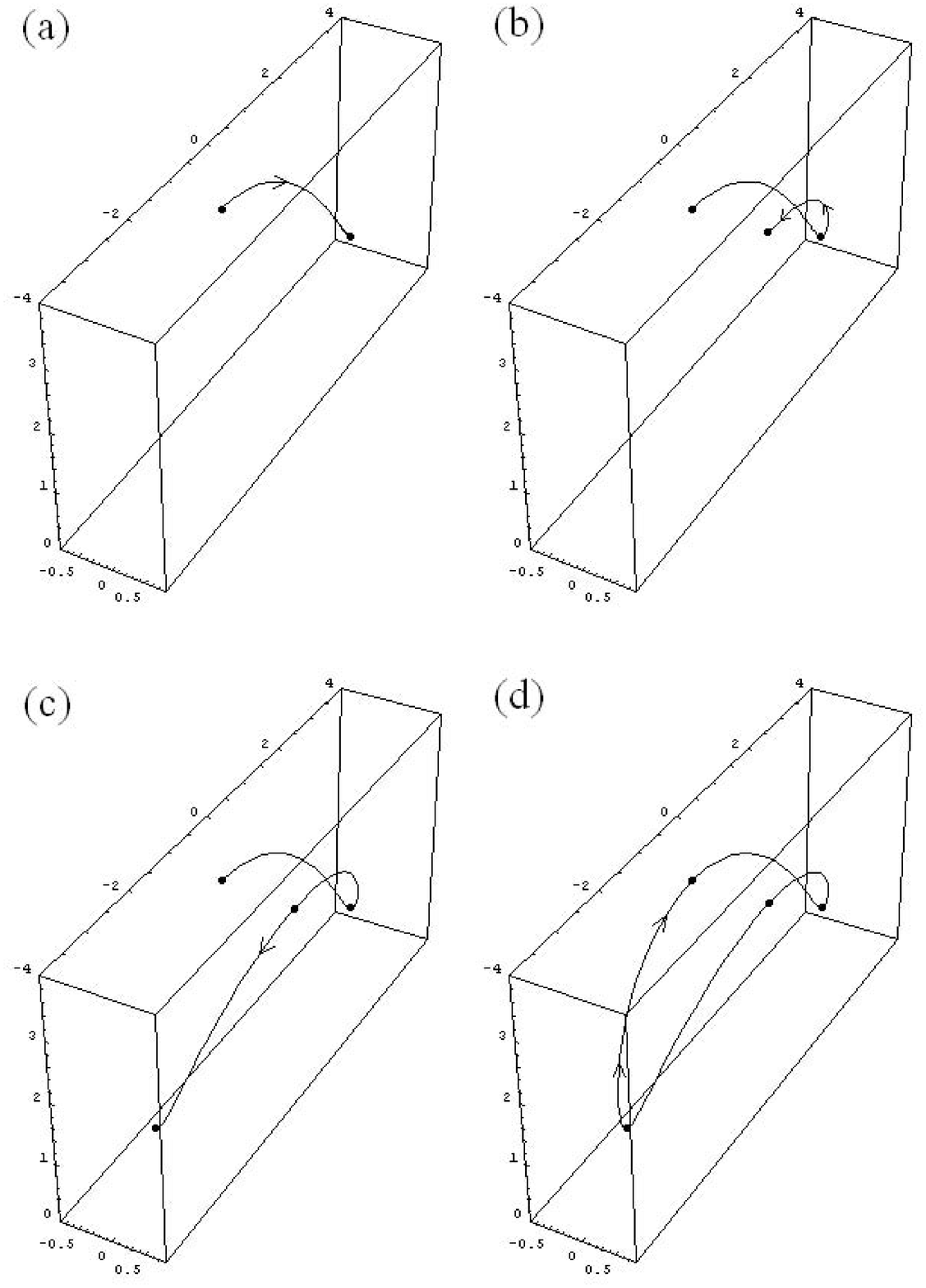}
\caption{Plots of $(\omega_{1},\omega_{2},\omega_{3})$ at different times: (a) $\tau = K$, (b) $\tau = 2K$, (c) $\tau = 
3K$, and (d) $\tau = 4K$.}
\label{fig:Rigid_Jac}
\end{figure}

When we substitute these expressions in the Euler equation (\ref{eq:euler_asym}) for $\omega_{2} \equiv \Omega_{2}\,y(\tau)$, we easily obtain the dimensionless Jacobi differential equation (\ref{eq:Jy_ode}), which can now be integrated, with the initial conditions $(\omega_{1}(0), \omega_{2}(0),
\omega_{3}(0)) = (-\,\Omega_{1}, 0, \Omega_{3})$, to yield 
\cite{Landau}
\begin{equation}
\left( \omega_{1},\; \omega_{2},\; \omega_{3} \right) \;=\; \left( -\;\Omega_{1}\,{\rm cn}\,\tau,\; \Omega_{2}\;
{\rm sn}\,\tau,\; \Omega_{3}\,{\rm dn}\,\tau \right).
\label{eq:omega_asym}
\end{equation}
These solutions are shown in Figure \ref{fig:Rigid_Jac} where the $4K$-periodicity is clearly observed. The solution of this problem is thus very elegantly expressed in terms of the Jacobi elliptic functions $({\rm sn}, {\rm cn}, {\rm dn})$. 

The separatrix solution ($m = 1$) corresponds to the case when $I_{0} = I_{2}$ for which $\Omega_{2} \equiv \Omega_{0}$, so that the separatrix solution is
\begin{eqnarray*}
\omega_{1}(\tau) & = & -\;\sqrt{\frac{I_{2}\,(I_{2} - I_{3})}{I_{1}\,(I_{1} - I_{3})}}\; \Omega_{0}\;{\rm sech}\,\tau, \\
\omega_{2}(\tau) & = & \Omega_{0}\;\tanh\,\tau, \\
\omega_{3}(\tau) & = & \sqrt{\frac{I_{2}\,(I_{1} - I_{2})}{I_{3}\,(I_{1} - I_{3})}}\; \Omega_{0}\;{\rm sech}\,\tau.
\end{eqnarray*}
Lastly, a symmetric top $(I_{1} = I_{2} \neq I_{3})$ corresponds to the limit $m \equiv 0$ (and $\Omega_{2} = \Omega_{1}$). The Jacobian solution 
(\ref{eq:omega_asym}) for a symmetric top therefore becomes $( \omega_{1},\; \omega_{2},\; \omega_{3} ) = ( -\;\Omega_{1}\,\cos\tau,\; \Omega_{1}\,\sin\tau,\; \Omega_{3} )$, where we have used the identities (\ref{eq:scd_0}) and $\tau \equiv (1 - I_{3}/I_{1})\,\Omega_{3}\,t$ is the dimensionless time.

The examples presented in this Section dealt with periodic orbital motions described in terms of Jacobi elliptic functions $({\rm sn}, {\rm cn}, {\rm dn})$ with arguments evaluated along the real axis (i.e., the two simple-pole singularities of the Jacobi elliptic functions are located on the imaginary axis of the Jacobi fundamental period-parallelogram). In the next Section, we consider the Weierstrass elliptic function, which has a double-pole singularity on the real axis of the fundamental period-parallelogram.

\section{\label{sec:Weierstrass_elliptic}Weierstrass Elliptic Functions}

The Weierstrass elliptic function $\wp(z; g_{2}, g_{3})$ is defined as the solution of the differential equation
\begin{eqnarray} 
(ds/dz)^{2} & = & 4\,s^{3} \;-\; g_{2}\,s \;-\; g_{3} \nonumber \\
 & \equiv & 4\;(s - e_{1})\,(s - e_{2})\;(s - e_{3}).
\label{eq:Weierstrass_inv}
\end{eqnarray}
Here, $(e_{1}, e_{2}, e_{3})$ denote the roots of the cubic polynomial $4s^{3} - g_{2}\,s - g_{3}$ (such that $e_{1} + e_{2} + e_{3} = 
0$), where the invariants $g_{2}$ and $g_{3}$ are defined in terms of the cubic roots as\cite{WW,HMF_Weiers}
\begin{equation}
\left. \begin{array}{rcl}
g_{2} & = & -4\,(e_{1}\,e_{2} + e_{2}\,e_{3} + e_{3}\,e_{1}) \\
 & = & 2\,\left( e_{1}^{2} + e_{2}^{2} + e_{3}^{2}\right) \\
 &  & \\
g_{3} & = & 4\;e_{1}\,e_{2}\,e_{3}
\end{array} \right\},
\label{eq:g_23}
\end{equation}
and $\Delta = g_{2}^{3} - 27\,g_{3}^{2}$ is the modular discriminant. Since physical values for the constants $g_{2}$ and $g_{3}$ are always real (and 
$g_{2} > 0$), then either all three roots are real or one root (say $e_{a}$) is real and we have a conjugate pair of complex roots $(e_{b}, e_{b}^{*})$ with ${\rm Re}(e_{b}) = -\,e_{a}/2$. The applications of Weierstrass elliptic functions are analyzed in terms of four different cases based on the signs of $(g_{3}, \Delta) = [(-,-), (-,+), (+,-), (+,+)]$, with two special cases $(g_{3} \neq 0, \Delta = 0)$ and $(g_{3} = 0,\Delta > 0)$.

\begin{figure}
\epsfysize=2in
\epsfbox{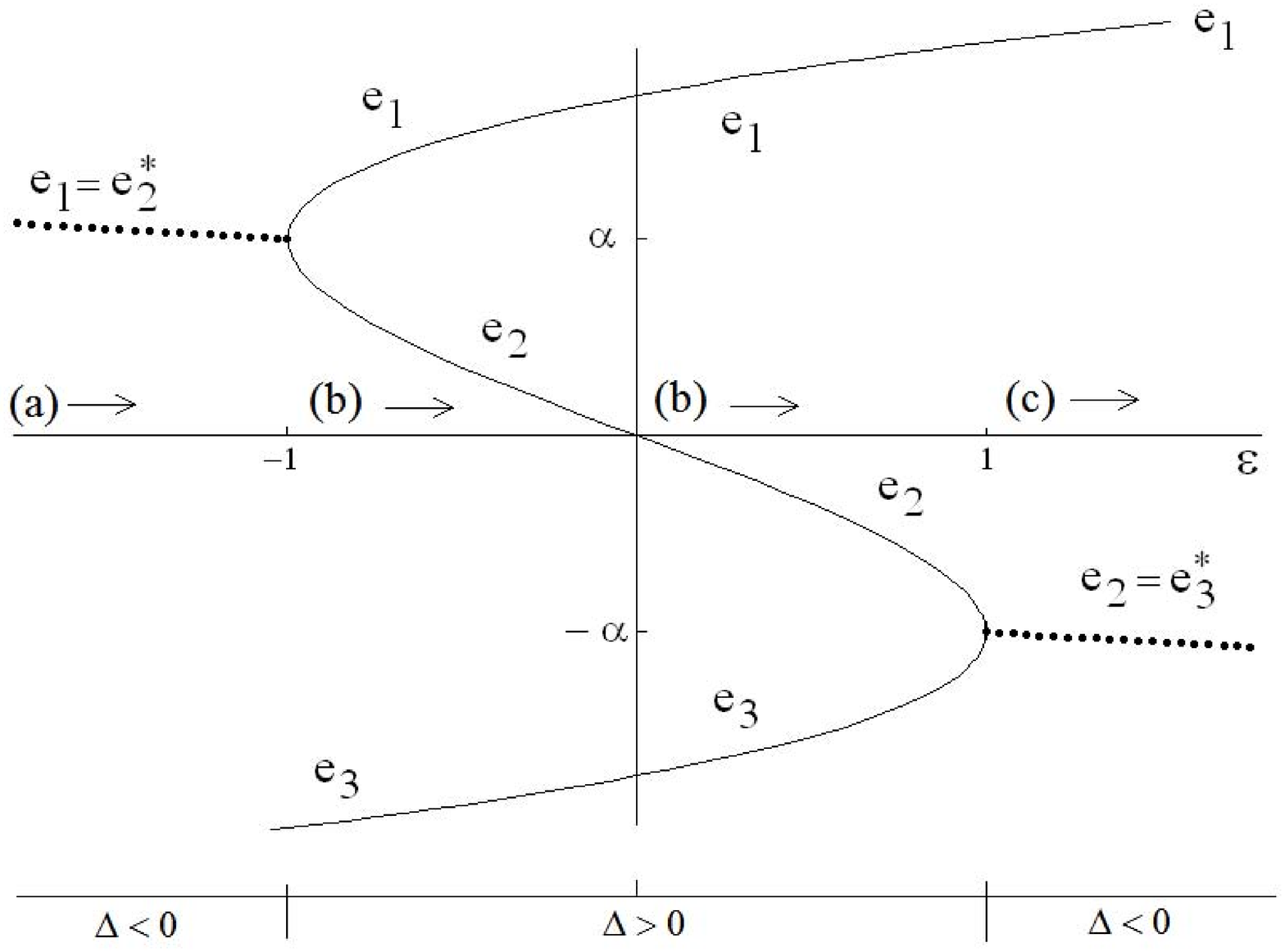}
\caption{Cubic roots $(e_{1}, e_{2}, e_{3})$ as a function of $\epsilon \equiv (3/g_{2})^{3/2}\,g_{3}$ with fixed value 
$g_{2}$, where $\alpha \equiv \sqrt{g_{2}/12}$ and $\Delta = g_{2}^{3}\,(1 - \epsilon^{2})$. The three roots satisfy 
$e_{1} + e_{2} + e_{3} = 0$. The regions (a)-(c) along the $\epsilon$-axis correspond to segments of the path shown in Fig.~\ref{fig:cubic_path}.}
\label{fig:cubic_root}
\end{figure}

\begin{figure}
\epsfysize=2in
\epsfbox{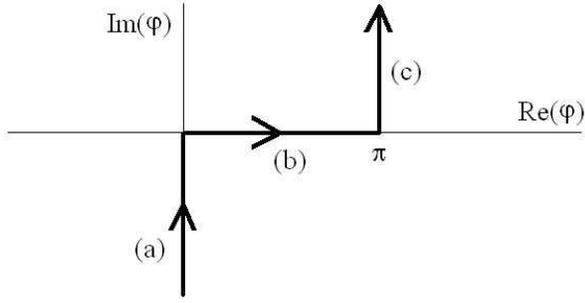}
\caption{Weierstrass path in complex $\varphi$ plane: (a) $\epsilon \leq -1$, (b) $-1 \leq \epsilon \leq 1$, and (c) $\epsilon \geq 1$.}
\label{fig:cubic_path}
\end{figure}

In general, the roots $(e_{1},e_{2},e_{3})$ of the cubic polynomial on the right side of Eq.~(\ref{eq:Weierstrass_inv}) can be expressed in terms of the parameters 
$\alpha \equiv \sqrt{g_{2}/12}$ and $\epsilon \equiv (3/g_{2})^{3/2}\,g_{3}$ as
\begin{widetext}
\begin{equation} 
\left( \begin{array}{c}
e_{1} \\
e_{2} \\
e_{3}
\end{array} \right) \;\equiv\; \alpha \left[\; \left( \epsilon + \sqrt{\epsilon^{2} - 1}\right)^{-1/3}
\left( \begin{array}{c}
1 \\
-\,e^{-i\pi/3} \\
-\,e^{i\pi/3}
\end{array} \right) \;+\; \left( \epsilon + \sqrt{\epsilon^{2} - 1}\right)^{1/3}
\left( \begin{array}{c}
1 \\
-\,e^{i\pi/3} \\
-\,e^{-i\pi/3}
\end{array} \right) \;\right],
\label{eq:cubic_123}
\end{equation}
\end{widetext}
and the discriminant is 
\begin{equation}
\Delta \;=\; g_{2}^{2} - 27\,g_{3}^{2} \;=\; g_{2}^{3}\,(1 - \epsilon^{2}). 
\label{eq:Delta_epsilon}
\end{equation}
These cubic roots are shown in Figure \ref{fig:cubic_root} as a function of $\epsilon$ for $\alpha = 1/2$ (i.e., $g_{2} = 3$); the polynomial $4s^{3} - g_{2}s - g_{3}$ is positive (and $ds/dz$ is real) to the left of the curve and negative (and $ds/dz$ is imaginary) to the right of the curve. The roots can be further parametrized by writing $\epsilon \equiv -\,\cos\varphi$, with $\epsilon + \sqrt{\epsilon^{2} - 1} = -\,\exp(-i\,\varphi)$ and 
\[ \left( \epsilon + \sqrt{\epsilon^{2} - 1}\right)^{\mp 1/3} \;=\; e^{\pm (\varphi - \pi)/3}, \] 
so that the three roots (\ref{eq:cubic_123}) become
\begin{equation} 
\left( \begin{array}{c}
e_{1} \\
e_{2} \\
e_{3}
\end{array} \right) \;\equiv\; 2\,\alpha\; \left( \begin{array}{c}
\cos[(\varphi - \pi)/3] \\
\cos[(\varphi + \pi)/3] \\
-\;\cos(\varphi/3)
\end{array} \right).
\label{eq:cubic_alphaphi}
\end{equation}
The three cubic roots are connected smoothly as shown in Figure \ref{fig:cubic_root} by following the path in the complex $\varphi$-plane shown in 
Fig.~\ref{fig:cubic_path}. Here, for $\epsilon \leq -1$, the imaginary phase $\varphi \equiv -i\,\psi$ (with $\psi \geq 0$) yields $e_{1} = a - i b = e_{2}^{*}$ (with $b > 0$) and $e_{3} = -\,2a < -1$; for $-1 \leq \epsilon \leq 1$, the real phase $\varphi \equiv \phi$ (with 
$0 \leq \phi \leq \pi$) yields $e_{1} > e_{2} > e_{3}$; and for $\epsilon \geq 1$, the complex phase $\varphi = \pi + i\,\psi$ (with $\psi \geq 0$) yields $e_{1} = 2a > 1$ and $e_{2} = -\,a - i b = e_{3}^{*}$ (with $b > 0$). Note that $e_{2} = 0$ (and $e_{1} = \sqrt{3}\,\alpha = -\,e_{3}$) for 
$g_{3} = 0$ (i.e., $\phi = \pi/2$); this case is called the {\it lemniscatic} case.\cite{HMF_Weiers} 

\begin{figure}
\epsfysize=2in
\epsfbox{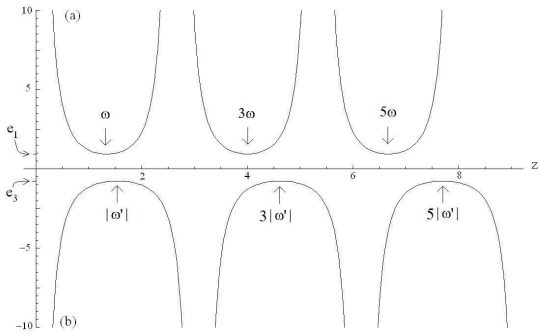}
\caption{Plots of (a) $\wp(z) > 0$ and (b) $\wp(i\,z) < 0$ for $g_{2} = 3$ and $g_{3} = 0.5$ (with $\epsilon = 0.5$) showing (a) the real period $2\,\omega$ and (b) the imaginary period $2\,\omega^{\prime}$ defined, respectively, by Eqs.~(\ref{eq:omega}) and (\ref{eq:omega_prime}).}
\label{fig:Weierstrass_fig}
\end{figure}

\begin{figure}
\epsfysize=2in
\epsfbox{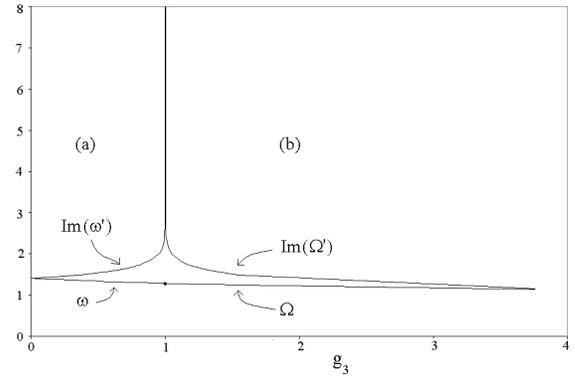}
\caption{Plots of (a) $\omega$ and ${\rm Im}(\omega^{\prime})$ for $0 < g_{3} < 1$ and $g_{2} = 3$ (i.e., $\Delta > 0$) and (b) $\Omega$ and ${\rm Im}(\Omega^{\prime})$ for $g_{3} > 1$ and $g_{2} = 3$ (i.e., $\Delta < 0$). Note that $\omega^{\prime}(g_{2},0) = i\,\omega(g_{2},0)$, $\Omega(g_{2}, 1) = \omega(g_{2}, 1)$, and both $(\Omega,\Omega^{\prime})$ decrease to zero as $g_{3}$ becomes infinite.}
\label{fig:omega}
\end{figure}

Figure \ref{fig:Weierstrass_fig} shows that, for $0 < \epsilon < 1$ (i.e., $\Delta > 0$), $\wp(z)$ has different periods $2\,\omega$ and 
$2\,\omega^{\prime}$ along the real and imaginary axes, respectively, with the half-periods $\omega$ and 
$\omega^{\prime}$ defined as
\begin{eqnarray} 
\omega(g_{2},g_{3}) & = & \int_{e_{1}}^{\infty}\;\frac{ds}{\sqrt{4 s^{3} - g_{2}\,s - g_{3}}}, \label{eq:omega} \\ 
\omega^{\prime}(g_{2},g_{3}) & = & i\;\int_{-\infty}^{e_{3}}\;\frac{ds}{\sqrt{|4 s^{3} - g_{2}\,s - g_{3}|}}. 
\label{eq:omega_prime}
\end{eqnarray}
The plots of $\omega(g_{2},g_{3})$ and $\omega^{\prime}(g_{2},g_{3})$ are shown in Fig.~\ref{fig:omega} for $g_{2} = 3$ as functions of $g_{3}$. Note that for $g_{3} = 0$ (with $e_{1} = -\,e_{3}$ and $e_{2} = 0$), we find that $\omega^{\prime} \equiv i\,\omega$ while $|\omega^{\prime}|$ approaches infinity as $g_{3}$ approaches one. Explicit calculations of the half periods $(\omega_{1}, \omega_{2}, \omega_{3})$ in terms of $\omega$ and 
$\omega^{\prime}$ are given in Appendix B.

For $\epsilon > 1$ (i.e., $\Delta < 0$), on the other hand, $\wp(z)$ has different periods $2\,\Omega$ and $2\,\Omega^{\prime}$ along the real and imaginary axes, respectively, with the half-periods $\Omega$ and $\Omega^{\prime}$ defined as
\begin{eqnarray} 
\Omega(g_{2},g_{3}) & = & \int_{e_{1}}^{\infty}\;\frac{ds}{\sqrt{4 s^{3} - g_{2}\,s - g_{3}}}, \label{eq:big_omega} \\
\Omega^{\prime}(g_{2},g_{3}) & = & i\;\int_{-\infty}^{e_{1}}\;\frac{ds}{\sqrt{|4 s^{3} - g_{2}\,s - g_{3}|}}. \label{eq:big_omega_prime}
\end{eqnarray}
The plots of $\Omega(g_{2},g_{3})$ and $\Omega^{\prime}(g_{2},g_{3})$ are shown in Fig.~\ref{fig:omega} for $g_{2} = 3$ as functions of $g_{3}$. Note that $\omega(g_{2},1) = \Omega(g_{2},1)$, $|\Omega^{\prime}|$ approaches infinity as 
$g_{3}$ approaches one, and that both $\Omega$ and $\Omega^{\prime}$ approach zero as $g_{3}$ approaches infinity.

\begin{figure}
\epsfysize=2in
\epsfbox{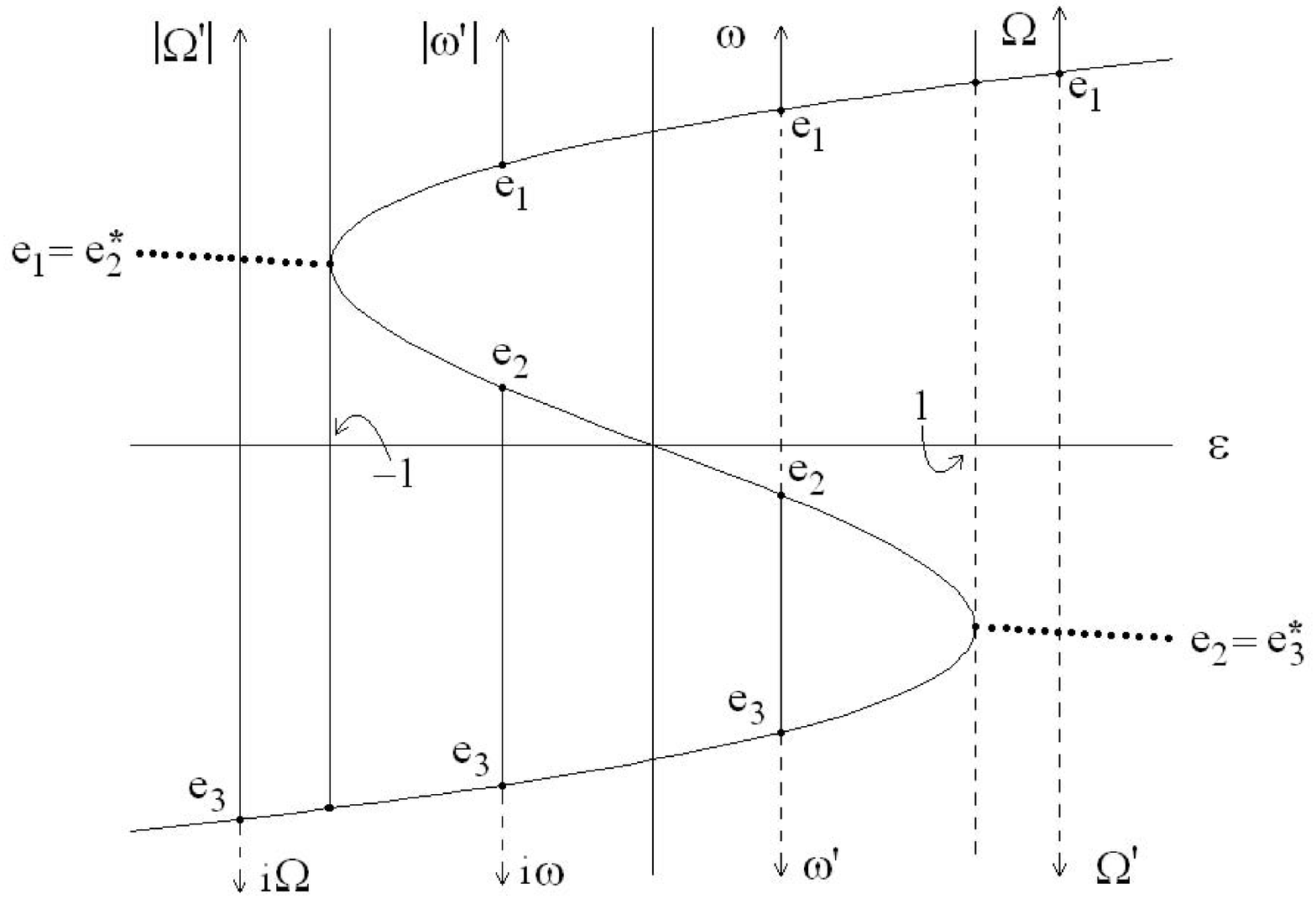}
\caption{Integration paths used to calculate the half-periods $(\omega,\, \omega^{\prime})$ for $\Delta > 0$ and $(\Omega,\,
\Omega^{\prime})$ for $\Delta < 0$. Solid (dashed) lines indicate where the polynomial $4 t^{3} - g_{2}\,t - g_{3}$ is positive (negative).}
\label{fig:Branch_period}
\end{figure}

Figure \ref{fig:Branch_period} shows the integration path used to calculate the half-periods $(\omega,\, \omega^{\prime})$ for $\Delta > 0$ and 
$(\Omega,\,\Omega^{\prime})$ for $\Delta < 0$. When the integration path includes a segment where the polynomial $4 t^{3} - g_{2}\,t - g_{3}$ is positive (solid line in Fig.~\ref{fig:Branch_period}), the contribution from this segment is a real-valued number. On the other hand, when the integration path includes a segment where the polynomial $4 t^{3} - g_{2}\,t - g_{3}$ is negative (dashed line in Fig.~\ref{fig:Branch_period}), the contribution from this segment is an imaginary number.

Table \ref{tab:root_HP} shows the cubic roots $e_{i} = (e_{1},e_{2},e_{3})$, defined by Eq.~(\ref{eq:cubic_alphaphi}), and the half periods $\omega_{i} = (\omega_{1},\omega_{2},\omega_{3})$, defined as\cite{WW}
\begin{eqnarray}
\omega_{i}(g_{2}, g_{3}) & \equiv & \int_{e_{i}}^{\infty}\; \frac{ds}{\sqrt{4 s^{3} - g_{2}\,s - g_{3}}} \nonumber \\
 & = & \int_{e_{i}}^{\infty}\; \frac{ds}{2\,\sqrt{(s - e_{1}) (s - e_{2}) (s - e_{3})}}.
\label{eq:omegai_def}
\end{eqnarray}
The cubic roots and half periods satisfy the following properties:
\begin{eqnarray}
\wp(\omega_{i}) & = & e_{i}, \label{eq:W_1} \\
\wp(z + \omega_{i}) & = & e_{i} \;+\; \frac{(e_{i} - e_{j})\,(e_{i} - e_{k})}{\wp(z) - e_{i}}, \label{eq:W_2} \\
\wp(z + 2\,\omega_{i}) & = & \wp(z), \label{eq:W_3}
\end{eqnarray}
where $i \neq j \neq k$ so that $\wp(\omega_{i} + \omega_{j}) = e_{k}$. 

\begin{figure}
\epsfysize=2in
\epsfbox{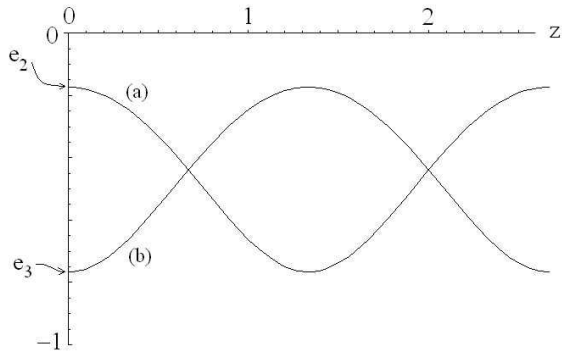}
\caption{Plots of (a) $\wp(z + \omega_{2})$ and (b) $\wp(z + \omega_{3})$ for $g_{2} = 3$ and $g_{3} = 0.5$ (with $\epsilon = 0.5$) over one complete period from $0$ to $2\,\omega_{1}$. Note that $\wp(\omega_{j}) = e_{j}$ for $j = 2$ or $3$ and $\wp(\omega_{i} + \omega_{j}) = e_{k}$, for $i = 1$ and $(j,k) = (2,3)$ or $(3,2)$.}
\label{fig:Weierstrass_per}
\end{figure}

Figure \ref{fig:Weierstrass_per} shows the plots of $\wp(z + \omega_{2})$ and $\wp(z + \omega_{3})$ for one complete period from $z = 0$ to $2\,\omega_{1}$, which clearly satisfies the identities (\ref{eq:W_1}) and (\ref{eq:W_3}). The singular behavior exhibited by Eq.~(\ref{eq:W_2}) at $z = \pm\,
\omega_{i}$, on the other hand, is shown in Fig.~\ref{fig:Weierstrass_fig} for $i = 1$ and $j, k \neq 1$.

\begin{widetext}

\begin{table}
\caption{\label{tab:root_HP}Cubic roots $(e_{1},e_{2},e_{3})$ and half periods $(\omega_{1},\omega_{2},\omega_{3})$ for $g_{2} = 3$ ($\alpha = 1/2$).
\cite{Footnote}} 
\begin{ruledtabular} 
\begin{tabular}{cccccccc}
$(g_{3}\;,\; \Delta)$ & $(-\;,\;-) $ & $(-\;,\; 0)$ & $(-\;,\; +)$ & $(0\;,\; +)$ & $(+\;,\; +)$ & $(+\;,\; 0)$ & $(+\;,\; -)$  \\ \hline
$e_{1}$ & $a - i\,b$ & $1/2$ & $d > 0$ & $\sqrt{3}/2$ & $c > 0$ & $2$ &  $2a > 1$ \\ \hline
$e_{2}$ & $a + i\,b$ & $1/2$ & $c - d > 0$ & $0$ & $d - c < 0$ & $-\,1$ &  $-\,a - i\,b$     \\ \hline
$e_{3}$ & $-\,2a < -1$ & $-\,1$ & $-\,c < 0$ & $-\,\sqrt{3}/2$ & $-\,d < 0$ & $-\,1$ &  $-\,a + i\,b$     \\ \hline
$\omega_{1}$ & $|\Omega^{\prime}| + i\,\Omega/2$ & $\infty$ & $|\omega^{\prime}|$ & $3^{-1/4}K(1/2)$ & $\omega$ & $\pi/\sqrt{6}$ &  $\Omega$     \\ \hline
$\omega_{2}$ & $-\,|\Omega^{\prime}| + i\,\Omega/2$ & $-\,\infty$ & $i\,\omega - |\omega^{\prime}|$ & $[3^{-1/4}K(1/2)](-1 + i)$ & $-\,\omega - 
\omega^{\prime}$ & $-\,i\,\infty$ & $-\,\Omega/2 - \Omega^{\prime}$ \\ \hline
$\omega_{3}$ & $-i\,\Omega$ & $-i\,\pi/\sqrt{6}$ & $-i\,\omega$ & $-i\,3^{-1/4}K(1/2)$ & $\omega^{\prime}$ & $i\,\infty$ & $-\,\Omega/2 + 
\Omega^{\prime}$ \\ 
\end{tabular}
\end{ruledtabular}
\end{table}

\end{widetext}

The Weierstrass elliptic function $\wp(z; g_{2}, g_{3})$ obeys the homogeneity relation \cite{WW,HMF_Weiers}
\begin{equation}
\wp\left(\lambda\,z; \lambda^{-4}\,g_{2}, \lambda^{-6}\,g_{3}\right) \;=\; \lambda^{-2}\;\wp(z; g_{2}, g_{3}),
\label{eq:W_hr}
\end{equation}
where $\lambda \neq 0$. By choosing $\lambda = -1$, for example, we readily verify that the Weierstrass elliptic function has even parity, i.e., 
$\wp(-z; g_{2}, g_{3}) = \wp(z; g_{2}, g_{3})$. On the other hand, for $\lambda = i$, we find that the half-period assignments for $g_{3} < 0$ in Table \ref{tab:root_HP} are based on the relation
\begin{equation} 
\wp(z; g_{2}, g_{3}) = -\;\wp(iz; g_{2}, |g_{3}|).
\label{eq:g3_negative}
\end{equation}
For example, for $-1 < g_{3} < 0$ (and $\Delta > 0$), we find for $\wp(\omega_{1}; g_{2}, g_{3})$:
\[ \wp(|\omega^{\prime}|; g_{2}, -\,|g_{3}|) \;=\; -\;\wp(\omega^{\prime}; g_{2}, |g_{3}|) \;=\; -\;(-\,d) \;=\; d, \]
which corresponds exactly to $e_{1} = d$ found in Table \ref{tab:root_HP} for the case $(g_{3}, \Delta) = (-, +)$. 

In general, the connections between the half-periods $(\omega_{1}^{+},\omega_{2}^{+},\omega_{3}^{+}) \rightarrow 
(\omega_{1}^{-},\omega_{2}^{-},\omega_{3}^{-})$ and the cubic roots $(e_{1}^{+},e_{2}^{+},e_{3}^{+}) \rightarrow (e_{1}^{-},e_{2}^{-},e_{3}^{-})$ as
$g_{3}$ changes sign from positive $(+)$ to negative $(-)$ are found in Table \ref{tab:root_HP} to be
\begin{equation}
\left. \begin{array}{rcl}
(\omega_{1}^{-},\omega_{2}^{-},\omega_{3}^{-}) & \equiv & (-i\,\omega_{3}^{+},
-i\,\omega_{2}^{+}, -i\,\omega_{1}^{+}) \\
 &  & \\
(e_{1}^{-},e_{2}^{-},e_{3}^{-}) & \equiv & (-\,e_{3}^{+},
-\,e_{2}^{+}, -\,e_{1}^{+})
\end{array} \right\}.
\label{eq:eomega_sign}
\end{equation}
Once again, these connections follow a non-standard convention. For example, according to the standard convention \cite{HMF_Weiers} for the case 
$(g_{3}, \Delta) = (+,-)$, the root $e_{2}$ is real (while $e_{1}^{*} = e_{3}$) and the corresponding half-period $\omega_{2}$ is also real (in contrast to the convention adopted in Table \ref{tab:root_HP}). The connections (\ref{eq:eomega_sign}) shown in Table \ref{tab:root_HP} are simply based on the smooth dependence of the cubic roots on the single parameter $\epsilon$ (for fixed $g_{2}$). These connections will enable us to describe consistent orbital dynamics in several problems in classical mechanics.

\subsection{Motion in a cubic potential}

\begin{figure}
\epsfysize=2in
\epsfbox{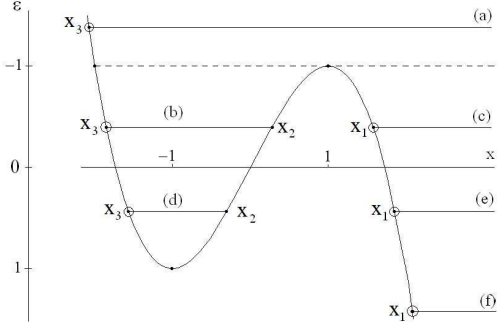}
\caption{Cubic-potential energy levels $E = x - x^{3}/3$ showing orbits (a) $E > 2/3$ (unbounded orbits; $\epsilon < -1$), (b) and (c) $0 < E < 2/3$ (bounded and unbounded orbits; $-1 < \epsilon < 0$), (d) and (e) $-2/3 < E < 0$ (bounded and unbounded orbits; $0 < \epsilon < 1$), and (f) $E \leq -2/3$ (unbounded orbit; $\epsilon \geq 1$). }
\label{fig:cubic_pot}
\end{figure}

As a first example of problems solved by the Weierstrass elliptic function $\wp$, we consider particle orbits in a (dimensionless) cubic potential $U(x) = x - x^{3}/3$. Here, the cubic-potential orbits $x(t)$ are solutions of the differential equation
\begin{eqnarray} 
\dot{x}^{2} & = & 2\, \left( E \;-\; x \;+\; \frac{x^{3}}{3} \right) \nonumber \\
 & \equiv & \frac{2}{3}\,(x - x_{1})\,(x - x_{2})\,(x - x_{3}),
\label{eq:Energy_cubic}
\end{eqnarray}
and the turning points $(x_{1},x_{2},x_{3})$ are shown in Figure \ref{fig:cubic_pot} (with $x_{1} + x_{2} + x_{3} = 0$). By writing $x(t) = 6\,s(t)$, Eq.~(\ref{eq:Energy_cubic}) is transformed into the standard Weierstrass elliptic equation (\ref{eq:Weierstrass_inv}), where the invariants are $g_{2} = 1/3$ and $g_{3} = -\,E/18$, so that $\epsilon \equiv -\,3E/2$. Note that bounded orbits exist only for $-\,1 < \epsilon < 1$ (i.e., $\Delta > 0$). 

\begin{figure}
\epsfysize=2.5in
\epsfbox{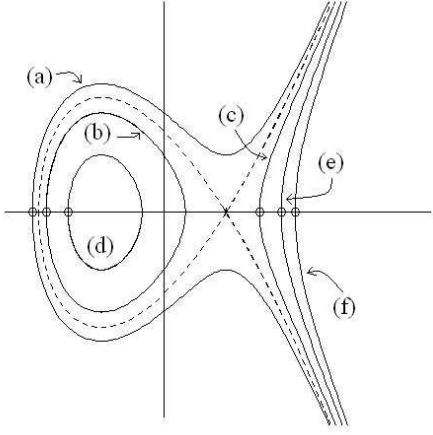}
\caption{Plots of $\dot{x}(t)$ versus $x(t)$ for cubic potential (\ref{eq:Energy_cubic}) shows bounded and unbounded orbits: Orbit (a) $E > 2/3$ ($\epsilon < -1$ and $\Delta < 0$); orbits (b)-(e) $-2/3 < E < 2/3$ ($-1 < \epsilon < 1$ and $\Delta > 0$); and orbit (f) $E < -2/3$ ($\epsilon > 1$ and $\Delta < 0$). The dotted lines are the bounded and unbounded separatrix orbits for $E = 2/3$ and circles denote particle positions at $t = 0$.}
\label{fig:Weierstrass_cubic}
\end{figure}

The cubic-potential solution is therefore $x(t) = 6\,\wp(t + \gamma)$, where the constant $\gamma$ is determined from the initial condition $x(0)$. In 
Fig.~\ref{fig:Weierstrass_cubic}, the orbits (a)-(f) are shown with initial conditions identified by a circle and a qualitative description of these orbits is summarized in Table \ref{tab:cubic}. Note that the turning points $x_{i} = 6\,e_{i}$ ($i = 1, 2, 3$) are simply related to the standard cubic roots $e_{i}$. Lastly, the separatrix solution is obtained from orbit (b) as $E$ approaches $2/3$ and the period $2\,|\omega^{\prime}|$ becomes infinite.

\begin{widetext}

\begin{table}
\caption{\label{tab:cubic}Bounded and unbounded orbits in a cubic potential (see Figs.~\ref{fig:cubic_pot} and \ref{fig:Weierstrass_cubic}).} 
\begin{ruledtabular} 
\begin{tabular}{cccccc}
Orbit & Energy & Time Range & Constant $\gamma$ & Period & Turning Point(s) \\ \hline
(a) & $E > 2/3$ & $-i\,\Omega < t < i\,\Omega$ & $-\,i\,\Omega$ & Unbounded & $e_{3} < 0$ \\ \hline
(b) & $0 < E < 2/3$ & $0 < t < 2\,|\omega^{\prime}|$ & $-\,i\,\omega$ & $2\,|\omega^{\prime}|$ & $e_{3} < e_{2}$ \\ \hline
(c) & $0 < E < 2/3$ & $-\,|\omega^{\prime}| < t < |\omega^{\prime}|$ & $|\omega^{\prime}|$ & Unbounded & $e_{1}$ \\ \hline
(d) & $-2/3 < E < 0$ & $0 < t < 2\,\omega$ & $\omega^{\prime}$ & $2\,\omega$ & $e_{3} < e_{2}$ \\ \hline
(e) & $-2/3 < E < 0$ & $-\,\omega < t < \omega$ & $\omega$ & Unbounded & $e_{1}$ \\ \hline
(f) & $E < -2/3$ & $-\,\Omega < t < \Omega$ & $\Omega$ & Unbounded & $e_{1}$ \\ 
\end{tabular}
\end{ruledtabular}
\end{table}

\end{widetext}

Note that the imaginary time range for orbit (a) takes into account the relation (\ref{eq:g3_negative}) since $g_{3} < 0$ for this orbit. In addition, the connections (\ref{eq:eomega_sign}) allow us to describe the orbits (a)-(f) in Figs.~\ref{fig:cubic_pot} and \ref{fig:Weierstrass_cubic} (and Table \ref{tab:cubic}) smoothly as the single (energy) parameter $\epsilon$ is varied. 

\subsection{\label{subsec:plan_w}Planar pendulum}

We now return to the planar pendulum problem of Sec.~\ref{subsec:ppend}, where we write $z = 1 \,-\,\cos\theta$ (i.e., $0 < z < 2$) and transform 
Eq.~(\ref{eq:ppendulum_eq}) into the cubic-potential equation
\begin{equation}
\left(z^{\prime}\right)^{2} \;=\; 2\,z\,(2 - z)\,(\epsilon - z),
\label{eq:ppend_W}
\end{equation}
with roots at $z = 0, 2$ and $\epsilon$. When $\epsilon < 2$, the motion is periodic between $z = 0$ and $z = \epsilon$, while the motion is periodic between $z = 0$ and $z = 2$ for $\epsilon > 2$. We recover the standard Weierstrass differential equation (\ref{eq:Weierstrass_inv}) by setting
\begin{equation}
z(\tau) \;=\; 2\,\wp(\tau + \gamma) \;+\; \mu,
\label{eq:u_W}
\end{equation}
where $\mu \equiv (\epsilon + 2)/3$ and the constant $\gamma$ is determined from the initial condition $z(0)$. 

\begin{figure}
\epsfysize=2in
\epsfbox{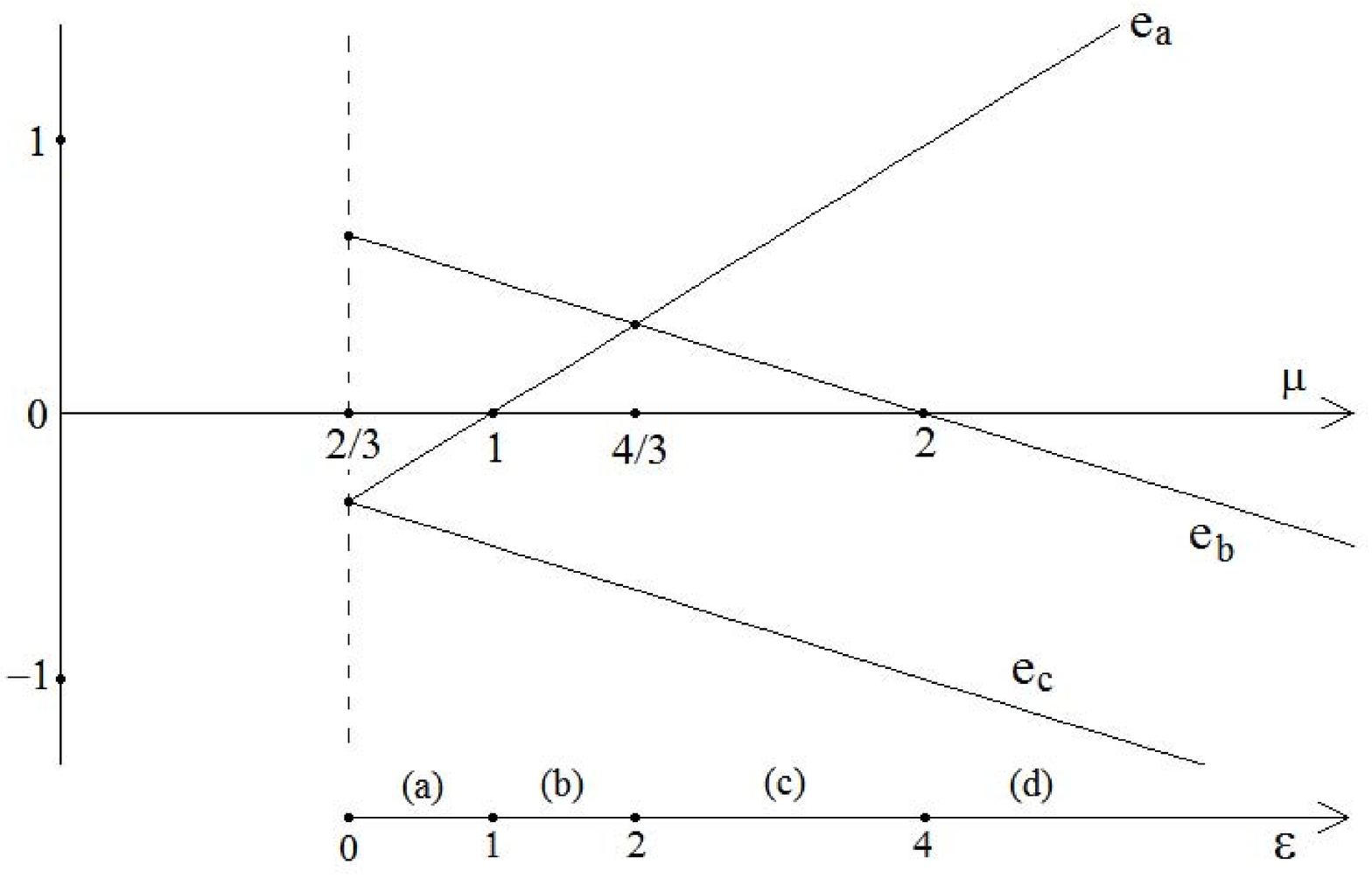}
\caption{Plots of the cubic roots $(e_{a}, e_{b}, e_{c})$ as functions of $\mu = (\epsilon + 2)/3$. The cases (a)-(d) are discussed in the text. Note that for cases (a) and (b), or $\epsilon < 2$, we find $e_{c} < e_{a} < e_{b}$, while for cases (c) and (d), or $\epsilon > 2$ we find $e_{c} < e_{b} < 
e_{a}$. The bounded motion of the planar pendulum $(-1 \leq z \leq 1)$ occurs between the two lowest cubic roots: $e_{c} < e_{a}$ (for $\epsilon < 2$) or $e_{c} < e_{b}$ (for $\epsilon > 2$).}
\label{fig:Weierstrass_pend}
\end{figure}

The root corresponding to $z = 0$ is labeled $e_{c} = -\,\mu/2$, the root corresponding to $z = 2$ is labeled $e_{b} = 1 - \mu/2$, and the root corresponding to $z = \epsilon$ is labeled $e_{a} = \mu - 1$ and we easily verify that $e_{a} + e_{b} + e_{c} = 0$ (see 
Fig.~\ref{fig:Weierstrass_pend}). The Weierstrass invariants $g_{2}$ and $g_{3}$ are
\[ g_{2} = 1 \;+\; 3\,(\mu - 1)^{2} \;\;{\rm and}\;\; g_{3} \;=\; \mu\,(\mu - 1)\,(\mu - 2), \]
and the modular discriminant is $\Delta = \epsilon^{2}\,(2 - \epsilon)^{2} \geq 0$. 

\begin{figure}
\epsfysize=2in
\epsfbox{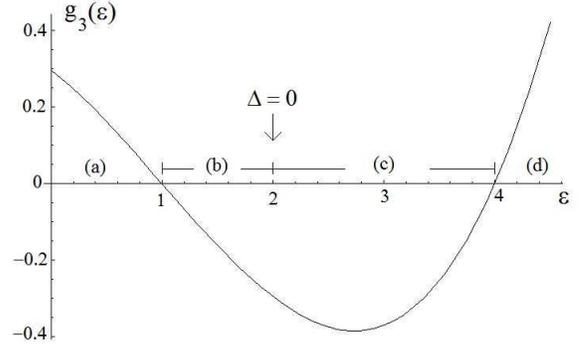}
\caption{Plot of the Weierstrass invariant $g_{3}$ as a function of $\epsilon$. For case (a), $g_{3} > 0$ and $\Delta > 0$; for cases (b) and (c), 
$g_{3} < 0$ and $\Delta \geq 0$; and for case (d) $g_{3} > 0$ and $\Delta > 0$.}
\label{fig:Wg3_pend}
\end{figure}

The planar pendulum is discussed in terms of 4 cases labeled (a)-(d) in Fig.~\ref{fig:Weierstrass_pend}. For cases (a) and (b), where $2/3 < \mu < 4/3$ (i.e., $0 < \epsilon < 2$), we find $e_{3} = -\,\mu/2 < e_{2} = \mu - 1 < e_{1} = 1 - \mu/2$, so that $\kappa = (e_{1} - e_{3})^{1/2} = 1$ and $m = (e_{2} - e_{3})/(e_{1} - e_{3}) = (3\mu - 2)/2 = \epsilon/2 < 1$. For cases (c) and (d), where $\mu > 4/3$ (i.e., $\epsilon > 2$), we find $e_{3} = -\,\mu/2 < e_{2} = 1 - \mu/2 < e_{1} = \mu - 1$, so that $\kappa = (e_{1} - e_{3})^{1/2} = (\epsilon/2)^{1/2}$ and $m = (e_{2} - e_{3})/(e_{1} - e_{3}) = 2/\epsilon < 1$. Figure \ref{fig:Wg3_pend} shows a plot of $g_{3}$ as a function of the parameter $\epsilon$, which can be used with the information presented in Table \ref{tab:root_HP} to describe the motion of the planar pendulum in terms of the Weierstrass elliptic function.

We first consider case (a), where $0 < \epsilon < 1$ (i.e., $2/3 < \mu < 1$ and $g_{3} > 0$), the periodic motion is bounded between $e_{3} = -\mu/2$ (i.e., $z = 0$) and $e_{2} = \mu - 1 < 0$ (i.e., $z = \epsilon$). Using the initial condition $z(0) = 0$, we find that $\wp(\gamma) = -\mu/2 \equiv e_{3}$ which implies that $\gamma = \omega^{\prime}$ (see $\omega_{3}$ in Table \ref{tab:root_HP} for $g_{3} > 0$ and $\Delta > 0$). The Weierstrass solution of the planar pendulum for $0 < \epsilon < 1$ is therefore
\begin{equation}
z(\tau) \;=\; 2\,\wp(\tau + \omega^{\prime}) \;+\; \mu, 
\label{eq:penlib_W1}
\end{equation}
with the period of oscillation $2\,\omega$. For the case (b), where $1 < \epsilon < 2$ (i.e., $1 < \mu < 4/3$ and $g_{3} < 0$), the periodic motion is bounded between $e_{3} = -\mu/2$ (i.e., $z = 0$) and $e_{2} = \mu - 1 < 0$ (i.e., $z = \epsilon$). Using the initial condition $z(0) = 0$, we find that $\wp(\gamma) = -\mu/2 \equiv e_{3}$ which implies that $\gamma = -\,i\,\omega$ (see $\omega_{3}$ in Table \ref{tab:root_HP} for $g_{3} < 0$ and $\Delta \geq 0$). The Weierstrass solution of the planar pendulum for $1 < \epsilon < 2$ is
\begin{equation}
z(\tau) \;=\; 2\,\wp(\tau - i\omega) \;+\; \mu,
\label{eq:penlib_W2}
\end{equation} 
with the period of oscillation $2\,|\omega^{\prime}|$. As expected, when $\epsilon \rightarrow 2$ (i.e., $\Delta \rightarrow 0$ and $m \rightarrow 
1$), the period $2\,|\omega^{\prime}|$ approaches infinity as we approach the separatrix. For case (c), where $2 < \epsilon < 4$ (i.e., $4/3 < \mu < 2$ and $g_{3} < 0$), the periodic motion is bounded between $e_{3} = -\mu/2$ (i.e., $z = 0$) and $e_{2} = 1 - \mu/2$ (i.e., $z = 2$), with the period of oscillation $2\,|\omega^{\prime}|$. Using the initial condition $z(0) = 0$, we find that $\wp(\gamma) = -\mu/2 \equiv e_{3}$ which implies that $\gamma = -\,i\,\omega$ (see $\omega_{3}$ in Table \ref{tab:root_HP} for $g_{3} < 0$ and $\Delta \geq 0$) and thus the Weierstrass solution of the planar pendulum for $2 < \epsilon < 4$ is again given by Eq.~(\ref{eq:penlib_W2}). Note that the separatrix solution $(\epsilon = 2$) is represented by orbits (b) and (c) as $|\omega^{\prime}| \rightarrow \infty$. Lastly, for case (d), where $\epsilon > 4$ (i.e., $\mu > 2$ and $g_{3} > 0$), the periodic motion is bounded between $e_{3} = -\mu/2$ (i.e., $z = 0$) and $e_{2} = 1 - \mu/2$ (i.e., $z = 2$). Using the initial condition $z(0) = 0$, we find that $\wp(\gamma) = -\mu/2 \equiv e_{3}$ which implies that $\gamma = \omega^{\prime}$ (see $\omega_{3}$ in Table \ref{tab:root_HP} for $g_{3} > 0$ and $\Delta > 0$) and thus the Weierstrass solution of the planar pendulum for $\epsilon > 4$ is again given by Eq.~(\ref{eq:penlib_W1}). 

We conclude our discussion of the planar pendulum by using the Jacobi and Weierstrass solution of this problem to establish a general relation between these elliptic functions. First, we use the Jacobi elliptic solution (\ref{eq:pen_lib}) for the case $\epsilon < 2$ and find
\begin{equation} 
z(\tau) \;=\; 2\;\sin^{2}\frac{\theta}{2} \;=\; 2\,m\;{\rm sn}^{2}(\kappa\,\tau\,|\,m),
\label{eq:penlib_J}
\end{equation}
where $m = \epsilon/2$ and $\kappa = 1$. By comparing Eqs.~(\ref{eq:penlib_W1}) and (\ref{eq:penlib_J}), we obtain the relation
\[ \wp(\tau + \omega_{3}) \;=\; -\,\frac{\mu}{2} \;+\; \left( \frac{3}{2}\,\mu - 1\right)\;{\rm sn}^{2}\left(\tau\,|\, \frac{3}{2}\,\mu - 1\right), \]
which is just one example of the general relation
\begin{equation}
\wp(\tau + \omega_{3}) \;\equiv\; e_{3} \;+\; (e_{2} - e_{3})\;{\rm sn}^{2}(\kappa\,\tau\,|\,m).
\label{eq:Psn_id}
\end{equation}
We note that this relation validates the relation (\ref{eq:omega1_K}) between the Weierstrassian half-period $\omega_{1} = \omega$ and the Jacobian quarter-period $K \equiv \kappa\,\omega$. Next, using the Jacobi elliptic solution (\ref{eq:pen_rot}) for $\epsilon > 2$, we find
\begin{eqnarray}
z(\tau) & = & 2\;{\rm sn}^{2}\left( \sqrt{\epsilon/2}\,\tau \;|\; 2/\epsilon \right) \nonumber \\
 & = & 2\;{\rm sn}^{2}\left( m^{1/2}\,\tau\;|\; m^{-1}\right),
\label{eq:penrot_W}
\end{eqnarray}
and thus we recover once again the relation (\ref{eq:Psn_id}).

\subsection{Spherical pendulum}

Our second physical example is the spherical pendulum (whose pendulum bob moves on the surface of a unit sphere), which is described here in terms of the cylindrical coordinates $(\rho, \varphi, 
z)$.\cite{Whittaker} The energy equation for a spherical pendulum of unit mass and unit length is
\[ \epsilon\,\nu^{2} \;=\; \frac{1}{2} \left( \dot{\rho}^{2} \;+\; \rho^{2}\,\dot{\varphi}^{2} \;+\; \dot{z}^{2} \right) \;+\; \nu^{2}\,z, \]
where $\epsilon$ denotes the normalized total energy ($\nu^{2} \equiv g$). By substituting $\rho(z) = \sqrt{1 - z^{2}}$ (where $-1 \leq z \leq 1$) and the angular momentum conservation law $\ell \equiv \rho^{2}\,\dot{\varphi}/\nu$, we obtain the differential equation
\begin{eqnarray}
(z^{\prime})^{2} & = & 2\,(\epsilon - z)\,(1 - z^{2}) \;-\; \ell^{2} \nonumber \\
 & \equiv & 2\,(z - z_{1})\,(z - z_{2})\,(z - z_{3}),
\label{eq:spen_ode}
\end{eqnarray}
where $z^{\prime}(\tau) \equiv \nu^{-1}\,\dot{z}$ and $z_{1} + z_{2} + z_{3} = \epsilon$. Because the right side of Eq.~(\ref{eq:spen_ode}) is negative at $z = \pm\,1$, the highest root $z_{1}\;(> 1 > z_{2} > z_{3} > -\,1)$ of the cubic polynomial is greater than 1 and is, therefore, unphysical (since the cylindrical radius $\rho$ then becomes imaginary). The periodic motion of the spherical pendulum is thus bounded between $z_{3} < z < z_{2}$. 

The differential equation (\ref{eq:spen_ode}) can be transformed into the standard differential equation (\ref{eq:Weierstrass_inv}) by setting $z(\tau)  = 2\,\wp(\tau + \gamma) + \mu$, where $\mu \equiv \epsilon/3$, the constant $\gamma$ is determined from the initial condition $z(0)$, and the invariants $g_{2}$ and $g_{3}$ are $g_{2} = 1 + 3\,\mu^{2}$ and $g_{3} = \ell^{2}/4 + \mu\;(\mu^{2} - 1)$. Note that, if $z(0) = z_{3} > -\,1$, then $\gamma = \omega_{3}$ (i.e., $\omega^{\prime}$ when $\Delta > 0$ and $\Omega^{\prime}$ when $\Delta < 0$) and the solution of the spherical pendulum problem for the $z$-coordinate is
\begin{equation}
z(\tau) \;=\; 2\,\wp(\tau + \omega_{3}) \;+\; \mu \;\equiv\; \sqrt{1 \;-\; \rho^{2}(\tau)}.
\label{eq:spend_sol}
\end{equation}
The motion is periodic with period $2\,\omega_{1}$ (i.e., $\omega$ when $\Delta > 0$ and $\Omega$ when $\Delta < 0$) on the unit circle in the $(\rho,z)$-plane. At the half-period $\omega_{1}$, we find $z(\omega_{1}) = 2\,\wp(\omega_{1} + \omega_{3}) + \mu = 2\,\wp(\omega_{2}) + \mu \equiv 
z_{2} < 1$ as expected.

The solution for the azimuthal angle $\varphi(\tau)$ is obtained from the angular-momentum conservation law $\varphi^{\prime}(\tau) = \ell/
\rho^{2}(\tau)$, which is integrated as \cite{Whittaker} 
\begin{eqnarray}
\varphi(\tau) & = & \ell\;\int_{0}^{\tau}\;\frac{ds}{1 - [2\,\wp(s + \omega_{3}) + \mu]^{2}} \label{eq:phi_spend0} \\
 & \equiv & -\;\frac{\ell}{4}\;\int_{0}^{\tau}\;\frac{ds}{[\wp(s + \omega_{3}) - \wp(\kappa)]\,[\wp(s + \omega_{3}) - \wp(\lambda)]}, \nonumber
\end{eqnarray}
where we used the initial condition $\varphi(0) = 0$ and the imaginary constants $\kappa$ and $\lambda$ are defined by the relations $\wp(\kappa) = 
-\,(1 + \mu)/2$ and $\wp(\lambda) = (1 - \mu)/2$ corresponding to $z = -\,1 < z_{3}$ and $z = +\,1 > z_{2}$, respectively. These constants also yield the relations $\wp^{\prime}(\kappa) = i\ell/2 = \wp^{\prime}(\lambda)$, obtained from the Weierstrass differential equation 
(\ref{eq:Weierstrass_inv}) for $z = \kappa$ and $\lambda$. These relations allow us to write 
Eq.~(\ref{eq:phi_spend0}) as
\begin{eqnarray}
\varphi(\tau) & = & \frac{i}{2}\;\int_{0}^{\tau} ds\; \left[\; \frac{\wp^{\prime}(\lambda)}{\wp(s + \omega_{3}) - \wp(\lambda)} \right. \nonumber \\
 &  &\left.\hspace*{0.8in}-\; \frac{\wp^{\prime}(\kappa)}{\wp(s + \omega_{3}) - \wp(\kappa)} \;\right],
\label{eq:phi_spend1}
\end{eqnarray}
where we used the identity $\wp(\lambda) - \wp(\kappa) = 1$. 

The integral (\ref{eq:phi_spend1}) can be solved exactly in terms of the quasi-periodic functions $\zeta(\tau)$ and $\sigma(\tau)$ associated with the Weierstrass elliptic function $\wp(\tau)$:\cite{WW} $\wp(\tau) \equiv -\,d\zeta(\tau)/d\tau$ and $\zeta(\tau) \equiv \sigma^{\prime}(\tau)/
\sigma(\tau)$. Using the identity
\begin{eqnarray*} 
\frac{\wp^{\prime}(\lambda)}{\wp(s) - \wp(\lambda)} & \equiv & \zeta(s - \lambda) \;-\; \zeta(s + \lambda) \;+\; 2\,\zeta(\lambda) \\
 & = & 2\,\zeta(\lambda) \;+\; \frac{d}{ds}\;\ln\left( \frac{\sigma(s - \lambda)}{\sigma(s + \lambda)} \right),
\end{eqnarray*}
we find the standard solution \cite{Whittaker} for the azimuthal motion of the spherical pendulum
\begin{widetext}
\begin{equation}
e^{2i\,\varphi(\tau)} \;=\; e^{2\tau\,[\zeta(\kappa) - \zeta(\lambda)]} \;\left[\; \left( \frac{\sigma(\tau + \omega_{3} + \lambda)\,
\sigma(\omega_{3} - \lambda)}{\sigma(\omega_{3} + \lambda)\,\sigma(\tau + \omega_{3} - \lambda)} \right) \;\times\; 
\left( \frac{\sigma(\tau + \omega_{3} - \kappa)\,\sigma(\omega_{3} + \kappa)}{\sigma(\omega_{3} - \kappa)\,\sigma(\tau + \omega_{3} + \kappa)} \right) \;\right],
\label{eq:phi_spend2}
\end{equation}
\end{widetext}
where we can easily verify that $\varphi(0) = 0$. This solution can be studied numerically with {\sf Mathematica} since its library of functions contains the Weierstrass elliptic function $\wp(\tau)$ and its associated quasi-periodic functions $\zeta(\tau)$ and $\sigma(\tau)$.

We now simplify Eq.~(\ref{eq:phi_spend2}) by recognizing that, since $-\,1 < z_{3} \leq \wp(\tau + \omega_{3}) \leq z_{2} < +\,1$ for real values of $\tau$, there must be imaginary numbers $i\,\alpha$ and $i\,\beta$ (where $\alpha$ and $\beta$ are real-valued constants) such that $\kappa 
\equiv \omega_{3} + i\,\alpha$ and $\lambda \equiv \omega_{3} + i\,\beta$. Using these substitutions, the solution (\ref{eq:phi_spend2}) becomes
\begin{widetext}
\begin{equation}
e^{2i\,\varphi(\tau)} \;=\; e^{2\tau\,[\zeta(\kappa) - \zeta(\lambda)]} \;\left[\; \left( \frac{\sigma(\tau + 2\,\omega_{3} + i\,\beta)\,
\sigma(-\,i\beta)}{\sigma(2\,\omega_{3} + i\,\beta)\,\sigma(\tau - i\,\beta)} \right) \;\times\; 
\left( \frac{\sigma(\tau - i\,\alpha)\,\sigma(2\,\omega_{3} + i\,\alpha)}{\sigma(-\,i\alpha)\,\sigma(\tau + 2\,\omega_{3} + i\,\alpha)} \right) 
\;\right].
\label{eq:phi_spend3}
\end{equation}
Next, using the identity\cite{WW} $\sigma(\tau + 2\,\omega_{3}) \equiv -\,\exp[2\,\eta_{3}\,(\tau + \omega_{3})]\; \sigma(\tau)$, where $\eta_{3} \equiv 
\zeta(\omega_{3})$, we now find
\[ \frac{\sigma(\tau + 2\,\omega_{3} + i\,\beta)}{\sigma(\tau - i\,\beta)} \;=\; -\; e^{2\,\eta_{3}\,(\tau + i\,\beta)}\; 
\frac{\sigma(\tau + i\,\beta)}{\sigma(\tau - i\,\beta)} \]
and
\[ \frac{\sigma(-\,i\beta)}{\sigma(2\,\omega_{3} + i\,\beta)} \;=\; -\; 
e^{-\,2i\,\eta_{3}\beta}\; \frac{\sigma(-\,i\,\beta)}{\sigma(i\,\beta)} \;\equiv\; e^{-\,2i\,\eta_{3}\beta}, \]
where we used the fact that $\sigma(\tau)$ is an odd function of $\tau$, so that $\sigma(-\,i\beta) = -\,\sigma(i\,\beta)$. Hence, we obtain
\[ \frac{\sigma(\tau + 2\,\omega_{3} + i\,\beta)\,\sigma(-\,i\beta)}{\sigma(2\,\omega_{3} + i\,\beta)\,\sigma(\tau - i\,\beta)} \;\equiv\;
-\;e^{2\,\eta_{3}\tau}\; \left[\; \frac{\sigma(\tau + i\,\beta)}{\sigma(\tau - i\,\beta)} \;\right], \]
and
\[ \frac{\sigma(\tau - i\,\alpha)\,\sigma(2\,\omega_{3} + i\,\alpha)}{\sigma(-\,i\alpha)\,\sigma(\tau + 2\,\omega_{3} + i\,\alpha)} \;\equiv\;
-\; e^{-2\,\eta_{3}\tau}\; \left[\; \frac{\sigma(\tau - i\,\alpha)}{\sigma(\tau + i\,\alpha)} \;\right]. \]
We combine these relations to obtain the simplified solution
\begin{equation}
e^{2i\,\varphi(\tau)} \;\equiv\; e^{2\tau\,[\zeta(\kappa) - \zeta(\lambda)]} \;\left[\; \frac{\sigma(\tau + i\,\beta)\,\sigma(\tau - i\,\alpha)}{\sigma(\tau - i\,\beta)\,\sigma(\tau + i\,\alpha)} \;\right].
\label{eq:phi_spend4}
\end{equation}
Lastly, we note that the ratio $\sigma(\tau + i\,\beta)/\sigma(\tau - i\,\beta)$ must have unit modulus for real $\tau$-values and so we may write
\begin{eqnarray*}
\ln \left( \frac{\sigma(\tau + i\,\beta)}{\sigma(\tau - i\,\beta)} \right) & = & i\;\int_{-\,\beta}^{\beta}\;
\zeta(\tau + i\,s)\;ds \\
 & = & 2i\;\int_{0}^{\beta}\;{\rm Re}\left[\zeta(\tau + i\,s)\right]\;ds.
\end{eqnarray*}
Hence, the azimuthal angle finally becomes
\begin{eqnarray}
\varphi(\tau) & = & \int_{\alpha}^{\beta}\;{\rm Re}[\zeta(\tau + i\,s)]\;ds \;+\; i\,\tau \left[\; \zeta(\omega_{3} + i\,\beta) \;-\; 
\zeta(\omega_{3} + i\,\alpha) \;\right] \nonumber \\
 & \equiv & {\rm Re} \left( \int_{\alpha}^{\beta} \left[ \zeta(\tau + i\,s) + \tau\,\wp(\omega_{3} + i\,s) \right]\,ds \right), 
\label{eq:phi_spend5}
\end{eqnarray}

\end{widetext}
which is now expressed only in terms of the quasi-periodic function $\zeta(\tau)$ and $\zeta^{\prime}(\tau) \equiv -\,\wp(\tau)$. After a full period 
$2\,\omega_{1}$, when the $(\rho,z)$-coordinates return to their initial values, the azimuthal angle has changed by an amount $\Delta\varphi \equiv 
\varphi(\tau + 2\,\omega_{1}) - \varphi(\tau)$ expressed as
\begin{equation} 
\Delta\varphi \;=\;  2\,\omega_{1} \;\int_{\alpha}^{\beta} \wp(\omega_{3} + i\,s)\;ds \;+\; 2\,\eta_{1}\,(\beta - \alpha), 
\label{eq:Delta_phi}
\end{equation}
where we used the identity\cite{WW} $\zeta(\tau + 2\omega_{1} + i\,s) = \zeta(\tau + i\,s) + 2\,\eta_{1}$ and $\eta_{1} \equiv \zeta(\omega_{1})$.

\subsection{Heavy symmetric top with one fixed point}

Our third physical example is provided by the motion of a symmetric top $(I_{1} = I_{2} \neq I_{3}$) with one fixed point described in terms of the energy equation 
\begin{eqnarray}
E & = & \frac{1}{2} \left[\; I_{1}\,\dot{\theta}^{2} \;+\; I_{3}\,\varpi_{3}^{2} \;+\; \frac{(p_{\varphi} - p_{\psi}\,\cos\theta)^{2}}{I_{1}\;\sin^{2}\theta} \;\right] \nonumber \\
 &  &+\; Mgh\,\cos\theta.
\label{eq:symtop}
\end{eqnarray}
where the total energy $E$ of the symmetric top (with total mass $M$ and principal moments of inertial $I_{1} = I_{2} \neq I_{3}$), $\varpi_{3}$ denotes the constant component of the angular velocity, and the angular momenta $p_{\varphi}$ and $p_{\psi}$ associated with the ignorable Eulerian angles $\varphi$ and $\psi$ are constants of motion. By defining the dimensionless parameters $\epsilon = (E - \frac{1}{2}\,I_{3}\,\varpi_{3}^{2})/(Mgh)$, $(a,b) = (p_{\varphi}/I_{1}\nu,\, p_{\psi}/I_{1}\nu)$, where $\nu^{2} = Mgh/(2\,I_{1})$, the differential equation for $u = \cos\theta$ is obtained from Eq.~(\ref{eq:symtop}) as
\begin{eqnarray}
(u^{\prime})^{2} & = & 4\,(1 - u^{2})(\epsilon - u) - (a - b\,u)^{2} \nonumber \\
 & = & 4\,(u - u_{1})\,(u - u_{2})\,(u - u_{3}),
\label{eq:tau_topfixed}
\end{eqnarray}
where a prime denotes a derivative with respect to the dimensionless time $\tau = \nu\,t$ and $u_{3} < u_{2} < u_{1}$ are the three roots of the cubic polynomial. Since the right side of Eq.~(\ref{eq:tau_topfixed}) is negative at $u = \pm\,1$, we conclude that $-1 < u_{3} < u_{2} < 1$ and $u_{1} > 1$ (which is unphysical for $u = \cos\theta$). The physical motion is therefore periodic in $\theta$ and is bounded between $u_{3} \equiv \cos\theta_{3}$ and $u_{2} \equiv \cos\theta_{2}$ (or $\theta_{2} < \theta(\tau) < \theta_{3}$). 

By using the change of integration variable $u = s + \mu$, where $\mu = (4\,\epsilon + b^{2})/12$, the differential equation (\ref{eq:tau_topfixed}) becomes the standard differential equation (\ref{eq:Weierstrass_inv}) for the Weierstrass elliptic function, with
\begin{eqnarray*}
g_{2} & = & 2\,\left(2 \;-\; ab \;+\; 6\,\mu^{2} \right), \\
g_{3} & = & a^{2} - 4\,\epsilon \;+\; 2\,\mu\;(2 \;-\; ab) \;+\; 8\,\mu^{3}.
\end{eqnarray*}
Hence, the solution is expressed in terms of the Weierstrass elliptic function as
\begin{equation} 
u(\tau) \;\equiv\; \cos\theta(\tau) \;=\; \wp(\tau + \gamma) \;+\; \mu,
\label{eq:W_heavy}
\end{equation}
where $\gamma$ is determined from the initial condition $\theta(0)$. Assuming that $-1 < u_{3} = e_{3} + \mu < u_{2} = e_{2} + \mu < 1 < u_{1} = e_{1} + \mu$, we choose $u(0) = e_{3} + \mu = u_{3}$ (i.e., $-1-\mu < e_{3} < 1 - \mu$) so that $\gamma \equiv \omega^{\prime}$ and, hence, at the half-period $\tau = \omega$, we find $u(\omega) = \wp(\omega + \omega^{\prime}) + \mu = e_{2} + \mu = u_{2}$. The solution for $\theta(\tau)$ is thus expressed as $\theta(\tau) = \cos^{-1}[ \wp(\tau + \omega^{\prime}) + \mu]$. 

Note that the sign of $\varphi^{\prime}$ depends on the sign of $a - b\,\cos\theta_{2} < a - b\,\cos\theta < a - b\,\cos\theta_{3}$. If $a > b\,\cos\theta_{2}$ (or $a < b\,\cos\theta_{3}$), $\varphi^{\prime}$ does not change sign as $\theta$ bounces between $\theta_{2}$ and $\theta_{3}$ and the motion in $\varphi$ involves monotonic azimuthal precession. If $a = b\,\cos\theta_{2}$ (or $a = b\,\cos\theta_{3}$), $\varphi^{\prime}$ vanishes at $\theta = \theta_{2}$ (or $\theta = \theta_{3}$) and the motion in $\varphi$ exhibits a cusp at that angle (since both $\theta^{\prime}$ and $\varphi^{\prime}$ vanish). If $a < b\,\cos\theta_{2}$, $\varphi^{\prime}$ vanishes at an angle $\theta_{2} < \theta_{0} < \theta_{3}$ and the motion in $\varphi$ exhibits retrograde motion between $\theta_{0} < \theta < \theta_{3}$.

The solution (\ref{eq:W_heavy}) for the Eulerian angle $\theta(\tau)$ can now be used to integrate the differential equations
\begin{eqnarray} 
\varphi^{\prime}(\tau) & = & \frac{a - b\,\cos\theta(\tau)}{1 \;-\; \cos^{2}\theta(\tau)} \label{eq:phi_heavy} \\
 & \equiv & \frac{i}{2} \left[\; \frac{\wp^{\prime}(\kappa)}{\wp(\tau + \omega_{3}) - \wp(\kappa)} \;-\; \frac{\wp^{\prime}(\lambda)}{\wp(\tau + \omega_{3}) - \wp(\lambda)} \;\right], \nonumber
\end{eqnarray}
and, defining $\psi^{\prime} \equiv (\varpi_{3}/\nu - b) + \chi^{\prime}$,
\begin{eqnarray} 
\chi^{\prime}(\tau) & = & \frac{b - a\,\cos\theta(\tau)}{1 \;-\; \cos^{2}\theta(\tau)}
\label{eq:psi_heavy} \\
 & \equiv & \frac{i}{2} \left[\; \frac{\wp^{\prime}(-\,\kappa)}{\wp(\tau + \omega_{3}) - \wp(-\,\kappa)} \;-\; \frac{\wp^{\prime}(\lambda)}{\wp(\tau + \omega_{3}) - \wp(\lambda)} \;\right], \nonumber
\end{eqnarray}
for the remaining Euler angles, where $\wp(\kappa) = 1 - \mu$ and $\wp(\lambda) = -\,(1 + \mu)$, with $\wp^{\prime}(\kappa) = i\,(a - b)$ and
$\wp^{\prime}(\lambda) = i\,(a + b)$. Note that since Eq.~(\ref{eq:phi_heavy}) is the same as Eq.~(\ref{eq:phi_spend1}), its solution is identical to Eq.~(\ref{eq:phi_spend5}) (even if the constants $\kappa$ and $\lambda$ are different). This same solution can also be applied to the solution for 
Eq.~(\ref{eq:psi_heavy}), where Eq.~(\ref{eq:phi_heavy}) is transformed into Eq.~(\ref{eq:psi_heavy}) by performing the change $(a,b) \rightarrow (b,a)$ and noting that $\wp(\tau)$ has even parity, i.e., $\wp(-\,\kappa) = \wp(\kappa)$, while $\wp^{\prime}(\tau)$ has odd parity, i.e.,
$\wp^{\prime}(-\,\kappa) = -\,\wp^{\prime}(\kappa)$.

\section{\label{sec:KdV}Other applications of elliptic functions in physics}

There are many more applications of elliptic functions in physics, including particle orbits and light paths in general relativity \cite{Gravity} and solutions of cosmological models.\cite{FRW} These applications unfortunately involve advanced topics that fall well outside the purpose of the present work. 

One interesting application worthy of discussion, however, deals with exact solutions of the Korteweg-de Vries (KdV) equation \cite{KdV_1}
\begin{equation} 
\pd{u}{t} \;+\; u\;\pd{u}{x} \;+\; \frac{\partial^{3}u}{\partial x^{3}} \;=\; 0
\label{eq:KdV_1}
\end{equation}
which describes the nonlinear evolution of the field $u(x,t)$. This nonlinear equation appears in many areas of physics \cite{KdV_2} and is a member of an important class of nonlinear partial differential equations that possesses soliton solutions.\cite{KdV_3,KdV_4,KdV_5}

A travelling-wave solution of the Korteweg-de Vries (KdV) equation (\ref{eq:KdV_1}) is a function of the form $u(x,t) = v(\xi)$, where $\xi = \kappa\,(x - c\,t)$ denotes the wave phase (with constants $\kappa$ and $c$ to be determined). Substituting this travelling-solution in the KdV equation, we obtain an ordinary differential equation for $v(\xi)$:
\[ (v \;-\; c)\;v^{\prime} \;+\; \kappa^{2}\;v^{\prime\prime\prime} \;=\; 0, \]
which can be integrated with respect to $\xi$ to yield
\begin{equation}
\kappa^{2}\;v^{\prime\prime} \;=\; \alpha \;+\; c\,v \;-\; \frac{1}{2}\,v^{2},
\label{eq:KdV_2}
\end{equation}
where $\alpha$ is a constant of integration. If we multiply Eq.~(\ref{eq:KdV_2}) with $v^{\prime}$ and integrate again with respect to $\xi$, we obtain
\begin{equation}
\frac{\kappa^{2}}{2}\;\left( v^{\prime}\right)^{2} \;=\; \left( \alpha\,v + \beta \right) \;+\; \frac{c}{2}\;v^{2} \;-\; \frac{1}{6}\,v^{3},
\label{eq:KdV_3}
\end{equation}
where $\beta$ is a second constant of integration. It is now immediately clear that $v(\xi) \equiv A\,\wp(\xi) + B$ can be expressed in terms of elliptic functions (where $A \equiv -\,12\,\kappa^{2}$, $B \equiv c$) because the right side of Eq.~(\ref{eq:KdV_3}) involves a cubic polynomial in $v$. The travelling-wave solution of the KdV equation 
(\ref{eq:KdV_1}) is
\[ u(x,t) \;=\; A\;\wp[\kappa\,(x - c\,t) \;+\; \gamma] + B, \]
where the constant $\gamma$ is determined from the initial condition $u(x,0) = u_{0}(x)$. 

Using the relation between the Weierstrass and Jacobi elliptic functions (see Appendix \ref{sec:W_J}), the travelling-wave solution to the KdV equation 
(\ref{eq:KdV_1}) may also be expressed as 
\cite{KdV_4}
\[ u(x,t) \;=\; a\;{\rm cn}^{2}[\kappa\,(x - c\,t) + \gamma\;|\; m] \;+\; b, \]
where $m = \sqrt{(r_{3} - r_{2})/(r_{3} - r_{1})}$, $a = r_{3} - r_{2}$, $b = r_{2}$, and $\kappa = \sqrt{(r_{3} - r_{1})/6}$; here, $r_{1} < r_{2} < 
r_{3}$ are the roots of the cubic polynomial on the right side of Eq.~(\ref{eq:KdV_3}). This second representation is known as the periodic 
{\it cnoidal}-wave solution of the KdV equation (\ref{eq:KdV_1}).

Lastly, we note that for the special case $\alpha = 0 = \beta$ in Eq.~(\ref{eq:KdV_3}), for which $r_{3} = 3\,c$ and $r_{1} = 0 = r_{2}$, then we find
$m = 1$, $a = 3\,c$, $b = 0$, $\kappa = \sqrt{c/2}$, and the travelling-wave solution becomes
\[ u(x,t) \;=\; 3\,c\;{\rm sech}^{2}\left[ \sqrt{\frac{c}{2}}\;(x - c\,t) \right], \]
which describes the well-known localized soliton solution of the KdV equation (\ref{eq:KdV_1}).

\section{\label{sec:summary}Summary}

We presented a brief introduction of the Jacobi and Weierstrass elliptic functions with applications in classical mechanics. The problem of the planar pendulum was used to establish a connection between the Jacobi and Weierstrass elliptic functions. The double periodicity of the elliptic functions was easily observed in each of the problems of classical mechanics discussed in Secs.~\ref{sec:Jacobi_elliptic} and \ref{sec:Weierstrass_elliptic}. We also briefly discussed applications of the elliptic functions in other areas of physics (e.g., travelling-wave solutions of the Korteweg-de Vries equation).

Because of their similarity with trigonometric functions $\sin z$ and $\cos z$, physicists are most familiar with the Jacobi elliptic functions 
${\rm sn}(z|m)$ and ${\rm cn}(z|m)$. This familiarity is further increased by the fact that the problems of the planar pendulum and the force-free asymmetric top are solved simply in terms of the Jacobi elliptic functions (as was demonstrated in Sec.~\ref{sec:Jacobi_elliptic}).

Physicists are usually less familiar with the Weierstrass elliptic function $\wp(z; g_{2}, g_{3})$, introduced in Sec.~\ref{sec:Weierstrass_elliptic}. The complexity of the Weierstrass elliptic function is partly due to the fact that the form of its fundamental period parallelogram depends on the invariants $g_{2}$ and $g_{3}$ (and the modular discriminant $\Delta = g_{2}^{3} - 27\,g_{3}^{2}$). This is in contrast to the fundamental period parallelogram of the Jacobi elliptic function which remains rectangular for all values of $m \neq 1$. The greatest advantage of the Weierstrass elliptic function, however, is that the function $\wp(z)$ itself and its derivative $\wp^{\prime}(z)$ can be used to represent any doubly-periodic function $f(z) 
\equiv A(\wp)\,\wp^{\prime} + B(\wp)$, where $A$ and $B$ are arbitrary functions. For example, using Eq.~(\ref{eq:Psn_id}), we find 
${\rm sn}^{2}(\kappa\,z|m) \equiv (e_{2} - e_{3})^{-1}\,[\wp(z; g_{2}, g_{3}) - e_{3}]$. This simplicity was further demonstrated by the solution of the planar pendulum in terms of a single expression $z(\tau) = 1 - \cos\theta(\tau) = 2\,\wp(\tau + \gamma) + \mu$ for all values of energy.

While the introduction of the Jacobi and Weierstrass elliptic functions in the undergraduate curriculum remains difficult, it is hoped that the present primer, combined with the easy access to mathematical software, will facilitate their dissemination.

\appendix

\section{\label{sec:W_J}Relation between Weierstrass and Jacobi elliptic functions}

In this Appendix, we explore the connection between the Jacobi and Weierstrass elliptic functions. For this purpose, we begin with the differential equation for the Weierstrass elliptic function
\begin{equation}
\left( \frac{dy}{dx}\right)^{2} \;=\; 4\,y^{3} \;-\; g_{2}\,y \;-\; g_{3},
\label{eq:W_ODE}
\end{equation}
and introduce the transformation 
\begin{equation}
y(x) \;=\; \alpha\;s^{p}(z) \;+\; \beta,
\label{eq:W_J}
\end{equation}
where $p$ is an integer ($\neq 0, 1$), the function $s(z)$ depends on the new coordinate $z \equiv \kappa\,x$, and $(\alpha, \beta, \kappa)$ are constants to be determined. Under the transformation 
(\ref{eq:W_J}), Eq.~(\ref{eq:W_ODE}) becomes
\begin{eqnarray}
\left( \frac{ds}{dz} \right)^{2} & = & \frac{4\alpha}{p^{2}\kappa^{2}}\;s^{2 + p} \;+\; \frac{12\beta}{p^{2}\kappa^{2}}\;s^{2} \nonumber \\
 &  &+\; \left( 12\,\beta^{2} \;-\; g_{2} \right) \frac{s^{2 - p}}{\alpha\,p^{2}\kappa^{2}} \nonumber \\
 &  &+\; \left( 4\,\beta^{3} \;-\; g_{2}\,\beta \;-\; g_{3}\right)\; \frac{s^{2 - 2p}}{\alpha^{2}p^{2}\kappa^{2}}.
\label{eq:J_ODE}
\end{eqnarray}
The constants $(\alpha, \beta, \kappa)$ and the integer $p \neq 0, 1$ are chosen such that the right side of Eq.~(\ref{eq:J_ODE}) has the Jacobi form 
$(1 - s^{2})\,(1 - m\,s^{2})$. For $p = 2$, we recover the Jacobi form if 
\begin{equation}
\beta \;=\; -\;(m + 1)\,\frac{\kappa^{2}}{3}
\label{eq:beta}
\end{equation}
is a root of the cubic polynomial $4\,\beta^{3} - g_{2}\,\beta - g_{3}$ (i.e., $\beta = e_{1}, e_{2},$ or $e_{3}$) and
\[ \alpha \;=\; m\;\kappa^{2} \;=\; \frac{12\,\beta^{2} - g_{2}}{4\,\kappa^{2}}. \]
For $p = -2$, on the other hand, we recover the Jacobi form if $\beta$, given by Eq.~(\ref{eq:beta}), is again a root of the cubic polynomial $4\,
\beta^{3} - g_{2}\,\beta - g_{3}$ (i.e., $\beta = e_{1}, e_{2},$ or $e_{3}$) and
\[ \alpha \;=\; \kappa^{2} \;=\; \frac{12\,\beta^{2} - g_{2}}{4\;m\,\kappa^{2}}. \]
Hence, the Jacobi elliptic function $s(z)$ is related to the Weierstrass elliptic function $y(x)$ in the case of $p = \pm\,2$.

An application of the first transformation $(p = 2)$ shows that for $m = (e_{2} - e_{3})/(e_{1} - e_{3})$ and $\kappa = \sqrt{e_{1} - e_{3}}$, we find $\alpha = e_{2} - e_{3}$ and $\beta = e_{3}$, and we obtain the relation
\begin{equation}
\wp(x + \omega_{2}; g_{2}, g_{3}) \;=\; e_{3} \;+\; (e_{2} - e_{3})\;{\rm cn}^{2}\left( \kappa\,x \;|\; m \right),
\label{eq:Pcn_2}
\end{equation}
or
\begin{equation}
\wp(x + \omega_{3}; g_{2}, g_{3}) \;=\; e_{3} \;+\; (e_{2} - e_{3})\;{\rm sn}^{2}\left( \kappa\,x \;|\; m \right),
\label{eq:Psn_3}
\end{equation}
which oscillate between $e_{2} = \wp(\omega_{2}) = \wp(\omega + \omega_{3})$ and $e_{3} = \wp(\omega + \omega_{2}) = 
\wp(\omega_{3})$ (see Fig.~\ref{fig:Weierstrass_per}), where $\omega = K(m)/\kappa$. Relation (\ref{eq:Psn_3}) plays a crucial role in expressing the Weiertrassian solution of the planar pendulum in terms of its Jacobian solution [see Eq.~(\ref{eq:Psn_id})].

An application of the second transformation $(p = -2)$ shows that for the same $m$ and $\kappa$ as the first transformation, we find $\alpha = e_{1} - e_{3}$ and $\beta = e_{3}$, and we obtain the relation
\begin{equation}
\wp(x; g_{2}, g_{3}) \;=\; e_{3} \;+\; \frac{e_{1} - e_{3}}{{\rm sn}^{2}\left( \kappa\,x \;|\; m \right)},
\label{eq:P_sn}
\end{equation}
which has singularities at $x = 0$ and $2\,K(m)/\kappa \equiv 2\,\omega$ and a minimum ($e_{1}$) at $x = K(m)/\kappa \equiv \omega$ (see upper plot in Fig.~\ref{fig:Weierstrass_fig}). We note that the relation (\ref{eq:P_sn}) is equivalent to the property (\ref{eq:W_2}) of the Weierstrass elliptic function:
\[ \wp(x) \;=\; e_{3} \;+\; \frac{(e_{1} - e_{3})\,(e_{2} - e_{3})}{\wp(x + \omega_{3}) - e_{3}}, \]
when substituting Eq.~(\ref{eq:Psn_3}) into the denominator of Eq.~(\ref{eq:P_sn}).

\section{\label{sec:math}Mathematical Details}

In this Appendix, we present mathematical details that relate the half-periods (\ref{eq:omegai_def}) of the Weierstrass elliptic function with the quarter periods $K$ and $K^{\prime}$ of the Jacobi elliptic function. We assume throughout this Appendix that $g_{3} > 0$ and $\Delta > 0$ (i.e., $0 < \epsilon < 1$). 

First, we start with $\omega_{1}(g_{2},g_{3})$, where $s > e_{1} > e_{2} > e_{3}$ in the integral (\ref{eq:omegai_def}). By introducing the change of variable \cite{JMP} $s = e_{3} + (e_{1} - e_{3})\,{\rm csc}^{2}\psi$, we readily obtain
\begin{eqnarray}
\omega_{1}(g_{2},g_{3}) & = & \int_{0}^{\pi/2}\; \frac{d\psi}{\sqrt{(e_{1} - e_{3}) - (e_{2} - e_{3})\,\sin^{2}\psi}}
\nonumber \\
 & \equiv & \frac{K(m)}{\sqrt{e_{1} - e_{3}}} \;=\; \omega(g_{2}, g_{3}),
\label{eq:omega1_K}
\end{eqnarray}
where $\omega$ is defined in Eq.~(\ref{eq:omega}), the modulus $m(\phi)$ of the Jacobian quarter period is defined by the relation
\[ m(\phi) \;=\; \frac{e_{2} - e_{3}}{e_{1} - e_{3}} \;=\; \frac{\sin[(\pi - \phi)/3]}{\sin[(\pi + \phi)/3]}, \]
using the definitions (\ref{eq:cubic_alphaphi}), and $0 < \phi < \pi$ along the segment (b) in Fig.~\ref{fig:cubic_path}. 

Next, we look at $\omega_{3}(g_{2},g_{3})$, which can also be expressed as
\begin{eqnarray}
\omega_{3}(g_{2},g_{3}) & = & -\;\int_{-\infty}^{e_{3}} \frac{ds}{\sqrt{4 s^{3} - g_{2}\,s - g_{3}}} \label{eq:omega3_def} \\
 & \equiv & \frac{i}{2}\;\int_{-\infty}^{e_{3}}\;\frac{ds}{\sqrt{(e_{1} - s) (e_{2} - s) (e_{3} - s)}}. \nonumber
\end{eqnarray}
By introducing the change of variable $s = e_{1} - (e_{1} - e_{3})\,{\rm csc}^{2}\psi$, we readily obtain
\begin{eqnarray}
\omega_{3}(g_{2},g_{3}) & = & i\;\int_{0}^{\pi/2}\; \frac{d\psi}{\sqrt{(e_{1} - e_{3}) - (e_{1} - e_{2})\,\sin^{2}\psi}}
\nonumber \\
 & \equiv & \frac{i\,K(m^{\prime})}{\sqrt{e_{1} - e_{3}}} \;=\; \omega^{\prime}(g_{2},g_{3}),
\label{eq:omega3_K}
\end{eqnarray}
where $\omega^{\prime}$ is defined in Eq.~(\ref{eq:omega_prime}) and
\[ m^{\prime}(\phi) \;=\; 1 \;-\; m(\phi) \;=\; \frac{e_{1} - e_{2}}{e_{1} - e_{3}} \;=\;
\frac{\sin(\phi/3)}{\sin[(\pi + \phi)/3]}. \]

Lastly, we look at $\omega_{2}(g_{2},g_{3})$, which can be written as
\begin{equation} 
\omega_{2} \;\equiv\; \omega_{1} + \frac{i}{2}\,\int_{e_{2}}^{e_{1}} \frac{ds}{\sqrt{(e_{1} - s) (s - e_{2}) (s - e_{3})}}.
\label{eq:omega2_13}
\end{equation}
We now introduce the change of variable \cite{JMP}
\[ s^{\prime} \;=\; e_{2} \;-\; \frac{(e_{1} - e_{2})\,(e_{2} - e_{3})}{(s - e_{2})}, \]
and readily find that the second integral in Eq.~(\ref{eq:omega2_13}) is simply $\omega_{3}(g_{2},g_{3})$ so that
\begin{eqnarray}
\omega_{2}(g_{2}, g_{3}) & = & \omega_{1}(g_{2},g_{3}) \;+\; \omega_{3}(g_{2},g_{3}) \nonumber \\
 & \equiv & \omega + \omega^{\prime} \;=\; \frac{K + i\,K^{\prime}}{\sqrt{e_{1} - e_{3}}}.
\label{eq:omega2_K}
\end{eqnarray}


\begin{thebibliography}{99}

\bibitem{Whittaker} E.~T.~Whittaker, {\it A Treatise on the Analytical Dynamics of Particles and Rigid Bodies}, 4th ed. (Dover, New York, 1937).

\bibitem{Landau} L.~D.~Landau and E.~M.~Lifshitz, {\it Mechanics} (Elsevier, 1976).

\bibitem{Taylor} J.~R.~Taylor, {\it Classical Mechanics} (University Science Books, Sausalito, California, 2005).

\bibitem{Goldstein} H.~Goldstein, C.~Poole, and J.~Safko, {\it Classical Mechanics}, 3rd ed. (Addison-Wesley, San Francisco, 2002).

\bibitem{Greenhill} A.~G.~Greenhill, {\it The Applications of Elliptic Functions} (Dover, New York, 1959).

\bibitem{WW} E.~T.~Whittaker and G.~N.~Watson, {\it A Course of Modern Analysis}, 4th ed. (Cambridge University, London, 1963).

\bibitem{HMF_Jacobi} L.~M.~Milne-Thomson, {\it Jacobi Elliptic Functions and Theta Functions} in {\it Handbook of Mathematical Functions}, M.~Abramowitz and I.~A.~Stegun, eds. (Dover, New York, 1965) chap.~16.

\bibitem{HMF_Weiers} T.~H.~Southard, {\it Weierstrass Elliptic and Related Functions} in {\it Handbook of Mathematical Functions}, M.~Abramowitz and I.~A.~Stegun, eds. (Dover, New York, 1965) chap.~18.

\bibitem{Erdos_Seiffert} P.~Erd\"{o}s, Am.~J.~Phys.~{\bf 68}, 888 (2000).

\bibitem{Footnote} The reader should be warned that only the case $(g_{3}, \Delta) = (+, +)$ follows the standard convention.\cite{HMF_Weiers} The convention for the remaining cases $(g_{3}, \Delta) = [(-,-), (-,+), (+,-)]$ in Table \ref{tab:root_HP} are based on the output of {\sf Mathematica}, on which Eq.~(\ref{eq:cubic_alphaphi}) and the Weierstrass path in Fig.~\ref{fig:cubic_path} are based, and the convention adopted for $\omega_{2}$ is such that $\omega_{1} + \omega_{2} + \omega_{3} = 0$.\cite{Greenhill}

\bibitem{Gravity} J.~B.~Hartle, {\it Gravity: An Introduction to Einstein's General Relativity} (Addison-Wesley, San Francisco, 2003).

\bibitem{FRW} B.~Ryden, {\it Introduction to Cosmology} (Addison-Wesley, San Francisco, 2003).

\bibitem{KdV_1} J.~Gratton and R.~Delellis, Am.~J.~Phys.~{\bf 57}, 683 (1989).

\bibitem{KdV_2} S.~Giambo, P.~Pantano, and P.~Tucci, Am.~J.~Phys.~{\bf 52}, 238 (1984).

\bibitem{KdV_3} C.~S.~Gardner, J.~M.~Greene, M.~D.~Kruskal, and R.~M.~Miura, Phys.~Rev.~Lett.~{\bf 19}, 1095 (1967).

\bibitem{KdV_4} P.~J.~Olver, {\it Applications of Lie Groups to Differential Equations} (Springer-Verlag, New York, 1986), sec.~3.2.

\bibitem{KdV_5} A. Degasperis, Am.~J.~Phys.~{\bf 66}, 486 (1998).

\bibitem{JMP} C.~L.~Critchfield, J.~Math.~Phys.~{\bf 30}, 295 (1989).


\end{thebibliography}
\end{document}